\documentclass[lettersize,journal]{IEEEtran}
\usepackage{amsmath,amsfonts}
\usepackage[T1]{fontenc}
\usepackage{algorithmic}
\usepackage{array}
\usepackage[caption=false,font=normalsize,labelfont=sf,textfont=sf]{subfig}
\usepackage{textcomp}
\usepackage{stfloats}
\usepackage{url}
\usepackage{verbatim}
\usepackage{graphicx}
\def\BibTeX{{\rm B\kern-.05em{\sc i\kern-.025em b}\kern-.08em
    T\kern-.1667em\lower.7ex\hbox{E}\kern-.125emX}}
\usepackage{balance}
\usepackage{booktabs}
\usepackage{makecell}
\usepackage{enumitem}
\usepackage{multirow}
\usepackage{multicol}
\usepackage{tabularx}
\usepackage{xcolor}
\usepackage{caption}
\usepackage{titlesec}
\allowdisplaybreaks[4]

\begin{document}
\let\WriteBookmarks\relax
\def\floatpagepagefraction{1}
\def\textpagefraction{.001}
\newcommand{\revise}[1]{\textcolor{black}{#1}}
\newcommand{\new}[1]{\textcolor{black}{#1}}
\captionsetup[subfloat]{labelfont={rm,normalsize}, textfont={rm,footnotesize}}

\title{Iterative Learning Control with Mismatch Compensation for Residual Vibration Suppression in Delta Robots}
\author{Mingkun Wu, Alisa Rupenyan, \textit{Member, IEEE} and Burkhard Corves
\thanks{This work was supported in part by NCCR Automation, grant agreement 51NF40\textunderscore180545 from the Swiss National Science Foundation, in part by China Scholarship Council under Grant 202106250025. AR acknowledges funding support by the Joh. Jacob Rieter-Stiftung (\textit{Corresponding author:  Mingkun Wu})}
\thanks{Mingkun Wu and Burkhard Corves are with the Institute of Mechanism Theory, Machine Dynamics and Robotics, RWTH Aachen University, Aachen 52062, Germany. Mingkun Wu is also with ZHAW Centre for Artificial Intelligence, ZHAW Zürich University of Applied Sciences, Winterthur 8401, Switzerland
(e-mail: wu@igmr.rwth-aachen.de, corves@igmr.rwth-aachen.de)}
\thanks{Alisa Rupenyan is with ZHAW Centre for Artificial Intelligence, ZHAW Zürich University
of Applied Sciences, Winterthur 8401, Switzerland (e-mail: rupn@zhaw.ch)}}

\markboth{ }%
{Shell \MakeLowercase{\textit{et al.}}: A Sample Article Using IEEEtran.cls for IEEE Journals}
\maketitle
\begin{abstract}
Unwanted vibrations stemming from the energy-optimized design of Delta robots pose a challenge in their operation, especially with respect to precise reference tracking. To improve tracking accuracy,
this paper proposes an adaptive mismatch-compensated iterative learning controller based on input shaping techniques. 
We establish a dynamic model considering the electromechanical rigid-flexible coupling of the Delta robot, which integrates the permanent magnet synchronous motor. Using this model, we design an optimization-based input shaper, considering the natural frequency of the robot, which varies with the configuration. We proposed an iterative learning controller for the delta robot to improve tracking accuracy. Our iterative learning controller incorporates model mismatch where the mismatch approximated by a fuzzy logic structure.
The convergence property of the proposed controller is proved using a Barrier Composite Energy Function, providing a guarantee that the tracking errors along the iteration axis converge to zero. \new{Moreover, adaptive parameter update laws are designed to ensure convergence.} Finally, we perform a series of high-fidelity simulations of the Delta robot using Simscape to demonstrate the effectiveness of the proposed control strategy.
\end{abstract}
\def\abstractname{Note to Practitioners}
\begin{abstract}
This paper addresses the problems of residual vibration suppression, trajectory tracking and velocity constraints of Delta robots simultaneously. Residual vibration of Delta robots becomes increasingly severe due to light weight design requirements, particularly under high-speed and high-acceleration conditions, which poses a significant challenge to the working accuracy of Delta robots. In this paper, we design an optimal input shaper to suppress the residual vibration effectively while considering the dynamic characteristics varies with the configuration of Delta robot. Compared to passive vibration suppression methods, this input shaping method eliminates the need for extra sensors and materials so that it can contribute to cost reduction. Furthermore, ensuring that the Delta robot can accurately track the reference trajectory is essential for avoiding unexpected vibrations. Since Delta robots are primarily employed in repetitive pick-and-place tasks, in order to leverage historical information sufficiently, we propose an adaptive iterative learning controller (AILC), in which we introduce fuzzy logic structure (FLS) to approximate model mismatch such as the unknown dynamics caused by parameter uncertainty, the damping of motors, etc. Finally, we verify the effectiveness of the proposed control strategy by a high-fidelity multi-domain simulation, integrating the permanent magnet synchronous motor (PMSM), rigid links, and flexible links. In future works, we aim at implementing the proposed control strategy in the actual Delta robot prototype. 
\end{abstract}




\section{Introduction}
\IEEEPARstart{O}{ver} the past decades, pick-and-place parallel robots have been widely used in food, electronics, packaging and other industries, due to their high speed and high acceleration motion capability \cite{1206799, harandi2023adaptive}. As typical representatives, Delta robots possess three translational degrees of freedom \cite{wu2020vibration}. However, in order to reduce the weight of robots to achieve higher acceleration and decrease the energy costs, the links of the robots are made from lightweight materials, such as carbon fiber, which inevitably results in the deterioration of the residual vibration and eventually undermines the positioning accuracy of the robots. The occurrence of vibration makes it more difficult to achieve high performance trajectory tracking control and may cause instability in extreme circumstances. Therefore, designing an effective controller and suppressing the residual vibration are of great significance for the application of Delta robots as well as other parallel robots.

Generally speaking, the approaches of vibration suppression can be divided into two categories, passive vibration suppression approaches \cite{switonski2007application, kang2005vibration} and active vibration suppression approaches \cite{mohamed2005vibration,longvibration,zhang2016active}. The passive approaches utilize additional damping materials to increase the damping ratio of the system. Although these kinds of methods can reduce the vibration effectively, the use of extra materials increases the manufacturing costs. On the contrary, the active approaches involve designing a control system to achieve trajectory tracking and vibration suppression simultaneously. The active approaches can be classified into feedback methods and feedforward methods. The feedback methods have to acquire real time vibration signals to form the closed-loop controllers \cite{zhang2022vibration,he2021vibration}, which requires the installation of additional sensors to measure the vibration signals. The low-level controllers have to be changed and redesigned by the users, which may not be allowed for some commercial parallel robots. The feedforward methods do not change the existing low-level controllers but just modify the input reference trajectory based on the dynamic behavior of the systems, which are obviously easier for realization. Thus, controller design and residual vibration suppression can be addressed separately. Among the feedforward methods, the input shaping techniques are very common and useful tools to suppress the residual vibration \cite{xu2019optimal,park2006design,alhassan2023precise}. The principle of input shaping techniques is to modify the reference trajectory, which results in a deviation between the original reference trajectory and the modified trajectory. Therefore it may not be suitable for those task conditions where the reference trajectory is designed strictly and cannot be adjusted. However, for all high-speed parallel robots including Delta robots used to carry out pick-and-place operations, only start and end points are crucial.

In order to achieve satisfactory working accuracy, numerous control methods had been proposed for Delta robots and other parallel robots in the past decades, including sliding mode control \cite{jia2019second,yang2018continuous}, robust control \cite{castaneda2014robust,yang2019modified}, intelligent control \cite{vo2020neural,zhou2023fuzzy}, etc. However, all these mentioned control strategies only depend on the error information of the current operation period. For those applications where robots repeatedly perform the same tasks, the historical error information is a valuable resource to improve the control performance. Previous research has illustrated that the trajectory has a profound impact on the vibration of Delta robots. Therefore, it is crucial to ensure that the robot can track the reference trajectory to prevent unexpected vibration. Delta robots are used to perform the pick-and-place operation, which is repetitive along the fixed trajectory and within the same duration time. Under this circumstance, iterative learning control (ILC) \cite{minakais2019database, pereira2023self} as a suitable control approach to achieve a high performance and high accuracy trajectory tracking, has been widely employed in various fields, such as high-speed trains,  precision motion stages, etc \cite{yu2020adaptive, balta2021learning}. From the few existing applications of ILC on parallel / Delta robots, the restrictions of resetting condition and repetitive trajectory are overcome in the ILC proposed in \cite{boudjedir2019iterative}.
It has been further developed in \cite{boudjedir2021adaptive} to achieve an  adaptive robust proportional-derivatives ILC strategy in which a robust term is introduced to compensate the repetitive and nonrepetitive disturbance. \new{However, the above robust ILC methods for Delta robots require the norm of disturbance to satisfy certain relationships, which causes difficulties in their application. Moreover, the selection of control gains is complexly restricted due to the requirement of stability guarantee. Uncertainty compensation is in this case a suitable approach to address this problem, particularly when the uncertainty arises from model mismatch.}
\revise{A model mismatch compensation method is proposed in \cite{balta2021learning} for precision motion stages}, in which the Gaussian process regression is used to predict the unknown model mismatch, and the error is compensated using a repetitive control approach. 
 
 To improve modeling, the coupling between the mechanical and electrical systems, which profoundly influences the dynamic behavior of robots \cite{sun2017singular}, needs to be accounted for. Thus, for completeness, the model of PMSM
 should be integrated into the dynamic model of Delta robots. \revise{Furthermore}, in order to guarantee the working safety, the robots' velocity needs to be constrained. To introduce state constraints in the control problem, an adaptive neural network control algorithm is proposed in \cite{chen2020adaptive} for a wheeled mobile robot with velocity constraints, in which the barrier Lyapunov function (BLF) is introduced to guarantee the velocity constraints. \revise{The derivation of convergence guarantees for the BLF is inspired by \cite{chen2020adaptive} in this paper.} For parallel robots, especially for Delta robot, studying the effect of velocity constraints control is not common. The excessive speed not only requires more energy consumption but also increases the risks associated with the operation. Therefore, it is of great importance to take angular velocity constraints into account. 


In this paper, we propose a control strategy combining input shaping techniques and ILC for Delta robot with PMSM and angular velocity constraints. The main contributions of this work are listed as follows.
\begin{enumerate}[topsep=0pt]
    \item We establish an integrated dynamic coupling mathematical model of Delta robots, where the flexibility of joints and links, as well as the dynamics of PMSM are taken into account. Based on the established model, we design an optimal input shaper by two global optimization objectives, which can suppress the residual vibration effectively.
    \item According to the concept of the singular perturbation method (SPM), we propose an adaptive mismatch-compensated
    iterative learning controller (AMCILC) for the rigid-body motion coordinates of a Delta robot. The iterative learning method can effectively address the repetitive tasks of Delta robots. We introduce the FLS to approximate the model mismatch including the damping term of PMSM. By using two designed adaptive iterative update laws, we employ a Barrier Composite Energy Function (BCEF) to prove the convergence property of the tracking errors, in which the barrier Lyapunov function can ensure the velocity constraints to be satisfied.
    \item Based on Simscape's multi physical domain coupling function, we establish a high-fidelity simulation model of a Delta robot system to verify the performance of the proposed input shaper-based adaptive mismatch-compensated iterative learning controller (IS-AMCILC).
\end{enumerate}

The rest of this paper is summarized as follows. The coupling dynamic mathematical model of a Delta robot is established and the optimal input shaper is designed in Section \uppercase\expandafter{\romannumeral2}. Then, in section \uppercase\expandafter{\romannumeral3}, the design and stability proof of IS-AMCILC are conducted in which the velocity constraints are taken into account. A series of simulations is performed to verify the effectiveness of the proposed control strategy in Section \uppercase\expandafter{\romannumeral4}. Finally, conclusions are drawn in Section \uppercase\expandafter{\romannumeral5}.
\section{Dynamic Modeling and Input Shaper Design}

In this section, we firstly establish a dynamic coupling mathematical model of the Delta robot including rigid-body parts, flexible parts and PMSM. Then, we design an optimal input shaper based on the proposed two global optimization objectives.
\subsection{Dynamic Modeling of Mechanical Structures}
\begin{figure}[!tbp]
    \centering
    \begin{minipage}{0.48\linewidth}
        \centering
        \includegraphics[width=1\textwidth]{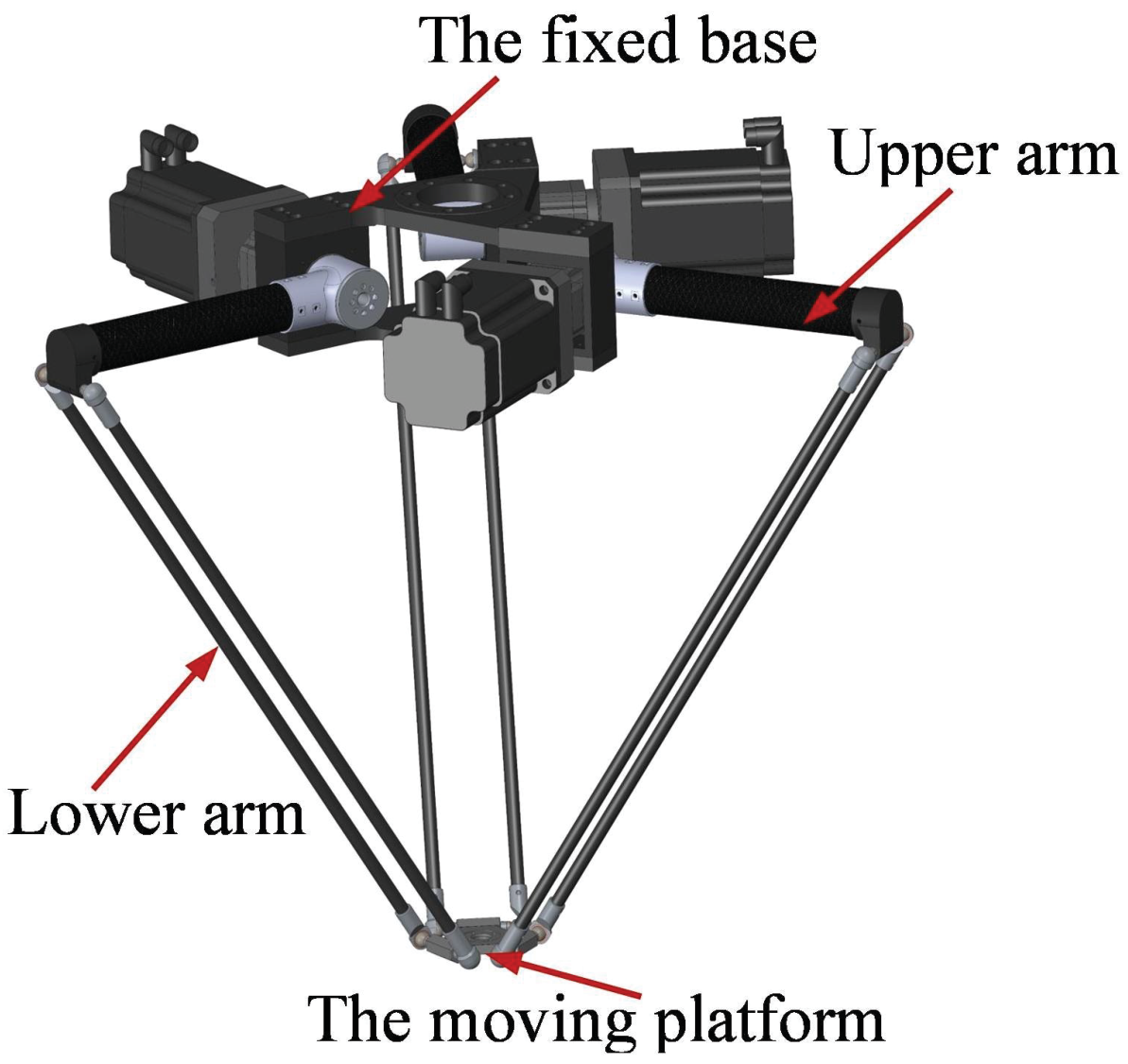}
        \caption{3D model of Delta robot.}
        \label{fig_1}
    \end{minipage}\hfill
    \begin{minipage}{0.48\linewidth}
        \centering
        \includegraphics[width=1\textwidth]{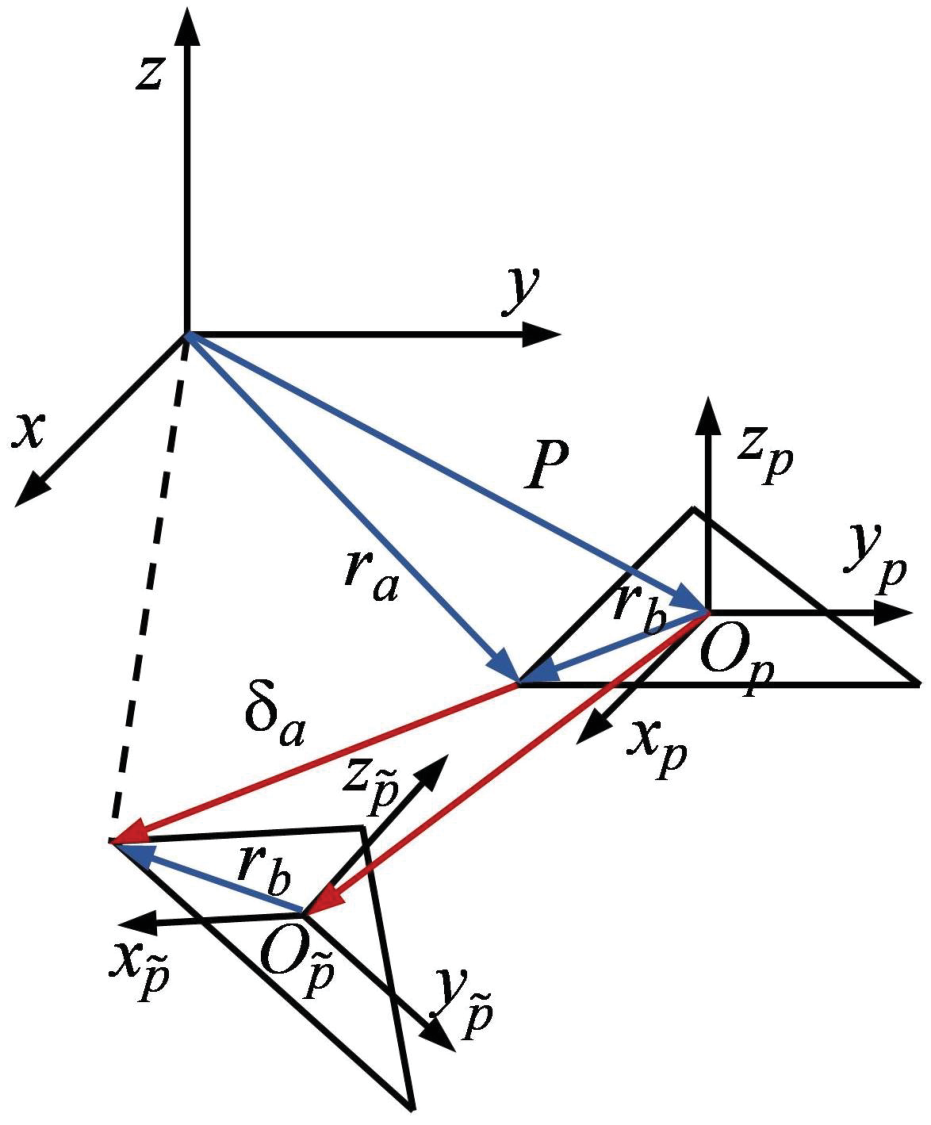}
        \caption{The deformation compatibility conditions.}
        \label{fig_2}
    \end{minipage}
\end{figure}
The 3D model of the Delta robot is illustrated in Fig.1. The robot consists of a fixed base, a moving platform (MP) and three identical kinematic chains, while each kinematic chain contains an upper arm and two lower arms. Before the dynamic modeling, it should be mentioned that according to the previous researches, the stiffness of the servo system and the gearbox play a critical role on the low modal of the Delta robot system \cite{shan2022residual}. Therefore, the connection between the gearbox and the upper arms will be regarded as a lumped spring system. Assuming the real input rotation angles, i.e., the output of the gearbox, can be denoted as $\theta \in \Re^{3\times1}$, we have
\begin{equation}
    \theta = \theta_{ri}+\theta_{fl},
\end{equation}
where $\theta_{ri} \in \Re^{3\times1}$ and $\theta_{fl} \in\Re^{3\times1}$ are the rotation angles of the rigid-body motion and flexible deformation, respectively.

In addition, the lower arms exhibit more noticeable flexibility than the upper arms and the moving platform in the practical operation. Therefore, only the lower arms will be regarded as flexible links in the dynamic modeling. Since the dynamic model is based on the floating frame of reference method, the real motion of a link can be expressed as the superposition of the rigid-body motion and the flexible deformation.

We utilize finite element method (FEM) to establish the flexible parts in this article. Considering that the length of the link is far larger than the radius, the Euler-Bernoulli beam theory is adopted. \new{Based on the principle of virtual work}, an unconstrained rigid-flexible coupling dynamic model of Delta robot can be obtained as
\revise{
\begin{align}
        F_u=&\sum_{i=1}^{9}H_i^T\left( \rho_i A_i\left( \int_{0}^{l_i}\Upsilon_id\bar{x}H_i\ddot{q_u} +\int_{0}^{l_i}\Theta_i^T\dot{\Theta}_i d\bar{x}H_i\dot{q_u} \right)\nonumber\right.\nonumber\\&\left.+\rho_i J_i\left(\int_{0}^{l_i}\Upsilon_{i,\theta}d\bar{x}H_i\ddot{q_u} +\int_{0}^{l_i}\Theta_{i,\theta}^T\dot\Theta_{i,\theta} d\bar{x}H_i\dot{q_u}\right)\right.\nonumber\\ &\left.+\int_{0}^{l_i}\Theta_i^T\rho_i A_ige_3d\bar{x} \right)+\sum_{i=1}^{9}K_{H,i}q_u+H_{\theta,p}^TI_{m}H_{\theta,p}\ddot q_u\nonumber\\&+ H_p^T\Theta_p^T\left( m\Theta_pH_p\ddot{q_u}+m\dot{\Theta}_pH_p\dot q_u+mge_z \right)\nonumber\\&+\sum_{i=1}^{3}H_{\theta}^T\left( m_{lump}\Upsilon_{lump}+m_{lump}\Theta_{lump}^T\dot\Theta_{lump} \right)H_{\theta}\nonumber\\&+\sum_{i=1}^{3}H_{\theta}^Tm_{lump}ge_3\nonumber\\=&M_u(q_u)\Ddot{q}_u+C_u(q_u,\dot q_u)\dot q_u +K_u(q_u)q_u+G_u(q_u),
\end{align}
where
\begin{align}
    \Upsilon_i=&\Theta_i^T\Theta_i,\ \Upsilon_{i,\theta}=\Theta_{i,\theta}^T\Theta_{i,\theta},\ e_3=[0,\ 0,\ 1]^T\nonumber\\ K_{H,i}=&H_i^Tk_iH_i,\ \Upsilon_{lump}=\Theta_{lump}^T\Theta_{lump}\nonumber
\end{align}}
where $ M_u(q_u)$, $C_u(q_u,\dot q_u)$, $K_u(q_u)$ and $G_u(q_u)$ are the inertia matrix, centrifugal force matrix, stiffness matrix and gravity vector, respectively. \revise{$\rho_i,\ A_i $ and $ J_i$ denote The density, the cross-sectional area, and the polar moment of inertia of the 
$i$th link, respectively, and $m$ and $I_m$ denote the mass and moment of inertia of MP, respectively. $g$ represents the gravitational acceleration, and $m_{lump}$ denotes the lumped mass between the upper arms and lower arms. $\Theta_i,\ \Theta_{i,\theta},\ \Theta_{lump}$ and $ \Theta_{p}$ denote the Jacobian matrices that map the generalized velocity $\dot q_f$ to the velocities of links, lumped mass and MP. $H_i,\ H_\theta$ and $H_p$ are Boolean indicated matrices \cite{liang2017nonlinear}. $k_i$ denotes the stiffness matrix of $i$th link} $q_u=[\theta_{ri}^T,\theta_{fl}^T,q_f^T]^T$, where $q_u$ is the generalized coordinate vector of the open-loop mechanism by cutting the joints. and $q_f$ is the generalized coordinate deformation vector of the flexible links including the small displacement of the moving platform caused by the flexible deformation of links. $F_u$ is the generalized force vector. \new{Note that the dimensions of $ M_u(q_u)$, $C_u(q_u,\dot q_u)$, $K_u(q_u)$, $G_u(q_u)$ and $F_u$ depend on the number of elements, e.g., we select only one beam element for each lower link in this paper, therefore, the dimensions of the above matrices would be $ M_u(q_u), C_u(q_u,\dot q_u), K_u(q_u) \in \Re^{48\times48}$ and $G_u(q_u), F_u \in \Re^{48\times1}$.}

The moving platform is connected to three identical kinematic chains, which implies that the flexible deformation of each link in the connection points is not independent but constrained by the closed-loop mechanism. Therefore, as shown in Fig. 2, we introduce the deformation compatibility conditions. As a result, in terms of the deformation compatibility conditions, the total number of the generalized coordinates are reduced to $q$, and their relationship can be obtained as
\begin{equation}
    q_u=T_{dc}q, \label{compatibility}
\end{equation}
where $T_{dc}\in \Re^{48\times30}$ is the deformation compatibility condition matrix. $q\in\Re^{30\times1}$ represents the generalized coordinate vector of the closed-loop mechanism. \new{The derivation of \eqref{compatibility} can be found in Appendix A.} Note that in order to facilitate the dynamic modeling process, we adopt the instantaneous substructure assumption \cite{zhao2021new}, which means that the mechanical structure is considered as a fixed structure at each infinitely small time interval $dt$, in other words the deformation compatibility condition matrix is viewed as a constant matrix, \new{i.e., $\dot q_u=T_{dc}\dot q$ and $\ddot q_u=T_{dc}\ddot q$.}

Finally, the rigid-flexible coupling dynamic model of the Delta robot with constraints can be deduced as
\begin{equation}
    M(q)\Ddot{q}+C(q,\dot q)\dot q+K(q)q+G(q)=F,
\end{equation}
where 
\begin{equation*}
\begin{split}
        M(q)=&T_{dc}^TM_u(q_u)T_{dc}, C(q,\dot q)=T_{dc}^TC_u(q_u,\dot q_u)T_{dc},\\K(q)=&T_{dc}^TK_u(q_u)T_{dc}, G(q)=T_{dc}^TG_u(q)T_{dc}, F=T_{dc}^TF_u.
\end{split}
\end{equation*}

\subsection{Dynamic Modeling of PMSM}
We first specify the simplifications and assumptions needed to develop the model. First, the effect of pulse width modulation is neglected. Some common assumptions that are widely used in the PMSM modeling are adopted here and the details can be found in \cite{sun2017singular}. Generally, the model of the PMSM is established in the $d$-$q$ axis based on the Clark and Park transformation. By means of the vector control technology, the electromagnetic torque of PMSM can be  deduced as 
\begin{equation}
    \tau_M=\frac{3}{2}p_M\psi_fi_q=K_ti_q,
\end{equation}
where $p_M$ is the number of pole pairs of the rotor. $\psi_f$ is the equivalent flux linkage of rotor magnetic field. $i_q$ is the q-axis stator current, and $K_t$ is the torque coefficient.

Accordingly, the mechanical equation of PMSM can be written as 
\begin{equation}
     \tau_M=I_M\frac{d\omega_M}{dt}+B\omega_M+\frac{\tau}{n_{gear}},
\end{equation}
where $I_M$ and $B$ are the moment of inertia and damping coefficient of PMSM, respectively. $n_{gear}$ is the gear ratio. $\omega_M$ is the angular velocity of the rotor. \new{$\tau$ denotes the driving torque vector of the active joint.}

Ultimately, the dynamic model of the mechanical structure and the PMSM will be integrated, and a completed dynamic model of Delta robot system can be obtained as
\begin{align}
\label{comple model}
    M(q)\Ddot{q}+C(q,\dot q)\dot q+K(q)q+G(q)=&F,\nonumber\\
    \bar I_M\frac{d\bar\omega_M}{dt}+\bar B\bar\omega_M+\frac{\bar\tau}{n_{gear}}=&\bar \tau_M.
\end{align}
where \new{$F=(\bar \tau,0,\dots,0)^T \in \Re^{30\times1}$ with $\bar\tau=[\tau_1,\tau_2,\tau_3]^T\in\Re^{3\times1}$, $\bar I_M = \text{diag}(\bar I_{M_1}, \bar I_{M_2}, \bar I_{M_3}$), $\bar B = \text{diag} (\bar B_1, \bar B_2, \bar B_3$) $\in\Re^{3\times3}$ and $ \bar\omega_M, \bar \tau_M \in\Re^{3\times1}$, as there are three motors in Delta robots.}
 Generally speaking, the damping term is difficult to be modeled. Therefore, a common practice is to generate the damping matrix by scaling the mass matrix and stiffness matrix, i.e. the Proportional damping. Hence, if the damping term is added, there will be an extra term $D(q)\dot q$ in \eqref{comple model}.
\subsection{Input Shaper Design}
Due to the orthogonality of the natural vibration modes, the following relationships can be obtained as
\begin{equation}
     M^{diag}=\Phi^TM\Phi, K^{diag}=\Phi^TK\Phi,
\end{equation}
where $\Phi=[\Phi_1, \Phi_2, \cdots,\Phi_n]$ is the modal matrix, and $\Phi_i$ is the modal vector of the $i$th mode. $ M^{diag}$ and $ K^{diag}$ are diagonal matrix. Then the natural angular frequency of the $i$th mode is calculated as 
$\omega_i=\sqrt{\frac{K^{diag}_i}{M^{diag}_i}}$, while $K^{diag}_i$ and $M^{diag}_i$ denote the $i$th element on the diagonal of matrix $K^{diag}$ and $M^{diag}$, respectively. Assuming $\zeta_i$ is the damping ratio of the $i$th mode. \new{For details about this part, we refer the reader to \cite{shan2022residual}.}

The residual vibration of the Delta robot system is determined by the superposition of the vibration modes. The vibration of the system is primarily dominated by the first few modes (\cite{shan2022residual}), especially the first mode. However, the first-order frequency varies depending on the configuration of the robot, which requires the designed input shaper to be robust enough so that it can achieve a better performance within the whole workspace. Compared with the ZV, ZVD and EI input shapers, an input shaper based on optimal control theory has been developed in (\cite{deng2015vibration}) with the following six design parameters, 
\begin{equation}
    \begin{split}
        A_1=\frac{1}{\Xi},A_2=-\frac{2\text{exp}(-\zeta_{d}\omega_{n}T)\text{cos}(\omega_{d}T)}{\Xi},\\
        A_3=\frac{\text{exp}(-2\zeta_{d}\omega_{n}T)}{\Xi},t_1=0,t_2=T,t_3=2T,
    \end{split}
\end{equation}
where $\Xi=1-2\exp(-\zeta_{d}\omega_{n}T)\cos{(\omega_{d}T)}+\exp(-2\zeta_{d}\omega_{n}T)$. $T=\frac{k_tT_d}{2}$ is the time lag of the impulse sequence with $0\le k_t \le 1$. $f_{n}$ and $\zeta_{d}$ are natural frequency and damping ratio respectively. $\omega_{d}=\omega_{n}\sqrt{1-\zeta_{d}^2}$, $\omega_n=2\pi f_n$ and $T_d=\frac{2\pi}{\omega_d}$.

 The input shaper proposed in \cite{deng2015vibration} exhibits a better robustness than ZV, ZVD and EI input shapers. Therefore, inspired by \cite{deng2015vibration}, we design an optimal input shaper for Delta robots to achieve a comprehensively satisfying performance within the entire workspace, in which the parameter selection of the input shaper will be encoded as an optimization problem. As an optimization objective, we adopt a widely used indicator to measure the performance of the input shapers, i.e., the percentage of residual vibration, which is the amplitude of the residual vibration after applying the last impulse. It can be written as
\begin{equation}
    V(\omega_n,\zeta_d,k_t)=\exp{(-\zeta_d\omega_nt_3)}\sqrt{C(\omega_n,\zeta_d)^2+S(\omega_n,\zeta_d)^2},
\end{equation}
where
\begin{equation*}
    \begin{split}
        C(\omega_n,\zeta_d)=\sum_{i=1}^{3}A_i\exp(\zeta_d\omega_nt_i)\cos(\omega_dt_i),\\
        S(\omega_n,\zeta_d)=\sum_{i=1}^{3}A_i\exp(\zeta_d\omega_nt_i)\sin(\omega_dt_i).
    \end{split}
\end{equation*}
\begin{figure}[!tbp]
    \centering
    \begin{minipage}{0.48\linewidth}
        \centering
        \includegraphics[width=1\textwidth]{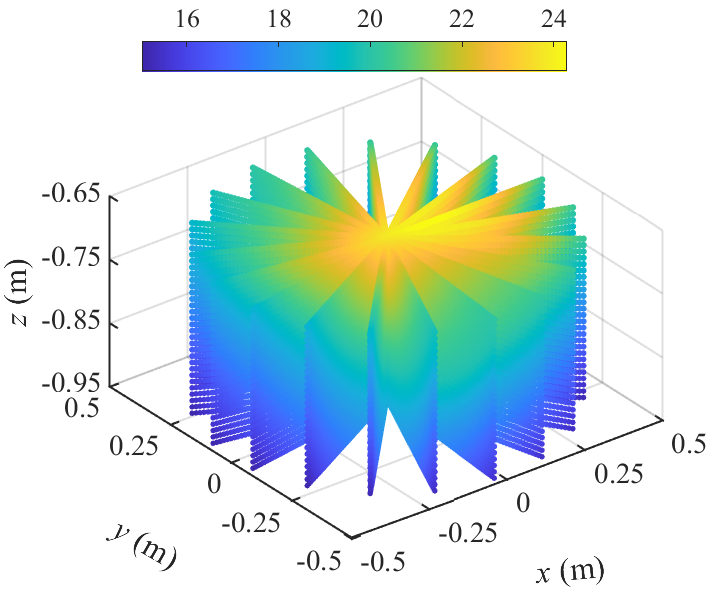}
        \caption{The first-order natural frequency of Delta robot within the whole workspace.}
        \label{fig_3}
    \end{minipage}\hfill
    \begin{minipage}{0.48\linewidth}
        \centering
        \includegraphics[width=1\textwidth]{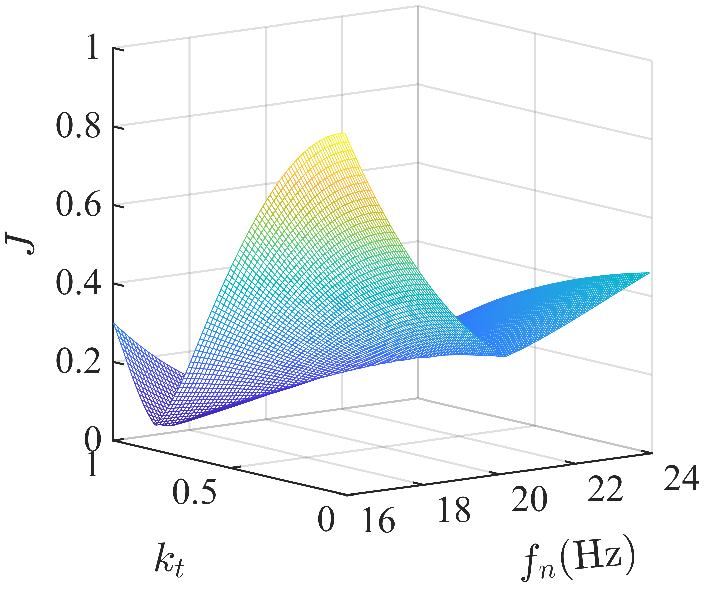}
        \caption{The trend of the optimization objective $J$ with respect to design variables.}
        \label{optimization}
    \end{minipage}
\end{figure}

Before the optimization, the optimizing variables and their ranges have to be determined. Firstly, although there exist three design variables $f_n$, $\zeta_d$ and $k_t$, the damping ratio has only a slight impact on the residual vibrations according to previous research results \cite{shan2022residual}. Therefore, only the natural frequency $f_n$ and the lag coefficient $k_t$ will be taken into account. The range of $k_t$ has been given above, i.e., $0 \le k_t \le 1$. The range of the $f_n$ can be determined by traversing the entire workspace. As shown in Fig. \ref{fig_3}, the maximum $f_n$ can be obtained on the top of the workspace, and it gradually decreases for lower work planes. Hence, the range of $f_n$ can be selected as $f_n \in [f_{min},f_{max}]$ in which $f_{min}=16 \text{Hz}$ and $f_{max}=24\text{Hz}$. For the optimization objectives, the maximum and average residual vibration percentage will be considered, separately. Defining the following two optimization objectives
\begin{equation}
    \begin{cases}
        J_1=\max V(\omega_n,\zeta_d=\zeta_{design},k_t),\\
        J_2=\frac{\int_{W}^{}V(\omega_n,\zeta_d=\zeta_{design},k_t)dW}{\int_{W}^{}dW}.
    \end{cases}
\end{equation}
\new{where $W$ denotes the workspace of Delta robots.}

In order to achieve a comprehensive residual vibration suppression performance, a weighted performance index is constructed.
\begin{equation}
    J=w_1J_1+w_2J_2,
\end{equation}
where $w_1>0$ and $w_2>0$, satisfying $w_1+w_2=1$, are two weight coefficients that can be determined manually. In this paper, these two coefficients are selected as: $w_1=w_2=0.5$.

In conclusion, we encode the input shaper design in the following optimization problems
\begin{align}
    \min \limits_{f_n,k_t}&\quad J(f_n,k_t) \nonumber\\
    s.t. &\quad f_n \in [f_{min},f_{max}]  \nonumber\\
    &\quad\  k_t \in [0,1]. \nonumber
\end{align}

Since the range of two optimization variables has been given, the optimal solutions can be found by grid search, where the grid sizes are set as 0.01 for two optimization variables. Based on \cite{shan2022residual}, the damping ratio will be selected as $\zeta_{design}=0.075$ here. The variation of the optimization objective with respect to the variables is presented in Fig. \ref{optimization}. It can be found that the optimal solution can be determined as $f_n=16.4$ Hz and $k_t=0.83$.

\section{Controller Design}
In this section, we design an AMCILC for Delta robot. First, we decompose the rigid-flexible coupling dynamic model of the Delta robot by SPM so that rigid-body motion and vibration can be addressed separately. Since the system exists the model mismatch caused by parameter uncertainty and unknown damping of PMSM, we utilize the FLS to approximate all aforementioned model mismatch. Then, we design an controller based on the concept of ILC to achieve high performance trajectory tracking, as ILC is an effective approach to deal with model mismatch \cite{balta2021learning}. 
\subsection{Model decomposing}
According to SPM, the rigid-flexible coupling dynamic model can be decomposed into two subsystems, i.e., a slow subsystem and a fast subsystem, where the slow subsystem is equivalent to the rigid-body dynamic model of the original system, and thus, can be utilized to design a controller to achieve the trajectory tracking. Therefore, inspired by SPM, we firstly deduce the rigid-body dynamic model with PMSM of the Delta robot system. Based on this model, we propose an adaptive fuzzy iterative learning controller.

The rigid-body dynamic model with PMSM of the Delta robot system at $k$th iteration can be expressed as
\begin{equation}
\label{rigid body}
    M_{r,k}(\theta_k)\ddot{\theta_k}+C_{r,k}(\theta_k,\dot\theta_k)\dot\theta_k+B_{r,k}\dot\theta_k+G_{r,k}(\theta_k)=u_k,
\end{equation}
with
\begin{align*}
    M_{r,k}(\theta_k)=&\frac{M_{rr,k}}{n_{gear}}+I_Mn_{gear},\quad B_{r,k}=Bn_{gear},\\
    C_{r,k}(\theta_k,\dot\theta_k)=&\frac{C_{rr,k}}{n_{gear}},\quad G_{r,k}(\theta_k)=\frac{G_{rr,k}}{n_{gear}},
\end{align*}
where $C_{r,k}(\theta_k,\dot\theta_k)=\bar{C}_{r,k}(\theta_k,\dot\theta_k)+\Delta C_{r,k}(\theta_k,\dot\theta_k)\in \Re^{3\times3}$, $M_{r,k}(\theta_k)=\bar{M}_{r,k}(\theta_k)+\Delta M_{r,k}(\theta_k) \in \Re^{3\times3}$ and $G_{r,k}(\theta_k)=\bar{G}_{r,k}(\theta_k)+\Delta G_{r,k}(\theta_k)\in \Re^{3\times1}$ are the inertia matrix, gyroscopic matrix and gravity vector at $k$th iteration, respectively.  $\bar{M}_{r,k}(\theta_k)$, $\bar{C}_{r,k}(\theta_k,\dot\theta_k)$ and $\bar{G}_{r,k}(\theta_k)$ are their nominal parts, while $\Delta M_{r,k}(\theta_k)$, $\Delta C_{r,k}(\theta_k,\dot\theta_k)$ and $\Delta G_{r,k}(\theta_k)$ are their unknown parts caused by parameter uncertainty, respectively. $B_{r,k}$ is the damping term of the PMSM, which is assumed to be unknown as well. These two parts are the main model mismatch we need to tackle in this paper. $u_k\in\Re^{3\times1}$ is the input vector at $k$th iteration. $\theta_k$ denotes the $\theta_{ri}$ at the $k$th iteration. \new{Hereafter, the subscript $k$ denotes the $k$th iteration for all variables.}

\subsection{Model mismatch approximation by FLS}
\textit{Lemma 1} \cite{wiktorowicz2015output,yu2023finite,shi2020adaptive}: For any unknown function $f(x)$ defined on a compact set $\Omega$, there exists a FLS $\vartheta^T\phi(x)$ such that,
\begin{equation}
    \underset{x\in\Omega}{\text{sup}}\left| f(x)-\vartheta^T\phi(x) \right|\le\sigma \quad\forall \sigma>0,
\end{equation}
where $x=[x_1,x_2,\cdots ,x_p]$ is the input of the fuzzy logic structure. $\vartheta=[\vartheta_1,\vartheta_2,\cdots,\vartheta_l]$ is the weight vector, and $\phi(x)=[\phi_1(x),\phi_2(x),\cdots,\phi_l(x)]$ is the fuzzy basis function vector with
\begin{equation}
   \phi_j(x)= \frac{\prod_{i=1}^{p}\mu_{F_{i}^j}(x_i)}{\sum_{j=1}^{l}\prod_{i=1}^{p}\mu_{F_{i}^j}(x_i)}.
\end{equation}

Then, according to Lemma 1, \eqref{rigid body} can be rewritten as
\begin{align}
\label{fuzzy dynamics}
    u_k=&M_{r,k}(\theta_k)\Ddot{\theta_k}+C_{r,k}(\theta_k,\dot\theta_k)\dot\theta_k+B_{r,k}\dot\theta_k+G_{r,k}(\theta_k)\nonumber\\
    =&\bar{M}_{r,k}(\theta_k)\Ddot{\theta_k}+\bar{C}_{r,k}(\theta_k,\dot\theta_k)\dot \theta_k+\bar{G}_{r,k}(\theta_k)\nonumber\\
    &-(\vartheta^T\phi(x_k)+\varepsilon),
\end{align}
where $\vartheta^T\phi(x_k)+\varepsilon=-(\Delta M_{r,k}(\theta_k)\Ddot{\theta_k}+\Delta C_{r,k}(\theta_k,\dot\theta_k)\dot \theta_k+\Delta G_{r,k}(\theta_k)+B_{r,k}\dot\theta_k)$ is the fuzzy structures used to approximate the model mismatch. $\vartheta=[\vartheta_1,\vartheta_2,\vartheta_3] \in \Re^{l\times3}$ is the unknown bounded weight vector, $\varepsilon=[\varepsilon_1,\varepsilon_2,\varepsilon_3]^T\in \Re^{3\times1}$ is the approximation error and satisfies $\varepsilon_i \le \bar \varepsilon_i$, where $\bar \varepsilon_i$ is an unknown constant. $x_k=[\theta_k^T, \dot \theta_k^T]$ denotes the input of the fuzzy structures.

Before designing the controller, we define the following error variables,
\begin{align}
    e_k=\theta_k-\theta_r,\\
    \dot e_k=\dot \theta_k-\dot\theta_r,
\end{align}
where $e_k \in \Re^{3\times1}$ and $\dot e_k \in \Re^{3\times1}$ are the angular position and velocity tracking errors at $k$th iteration, respectively. $\theta_r$ and $\dot\theta_r$ are the reference angular position and velocity, separately. 

Then, we introduce a combination of errors as follows,
\begin{equation}
\label{auxiliary variable}
    \eta_k=\dot e_k+\sigma e_k,
\end{equation}
where $\sigma$ is a positive constant.

To ensure that the constrained states do not exceed the predefined ranges, motivated by the BLF design method in (\cite{liu2019rbfnn}), we introduce a BLF,
\begin{equation}
\label{BLF}
    V_{b,i}=\frac{v_c^2}{\pi}\tan(\frac{\pi\eta_i^2}{2v_c^2}),
\end{equation}
where $v_c$ is a positive constant.

 It can be found from \eqref{BLF} that $V_{b,i}$ will tend to infinity, when the combination of errors $\eta_i$ close to $v_c$, which means if we want to prevent $\eta_i$ exceeding a specific range, the controller algorithm has to guarantee the BLF is bounded. Therefore, the boundedness of angular velocity can be ensured.

In order to enhance the operational safety and mitigate the vibration caused by excessive speed, it is of great importance to limit the angular velocity within a predefined range, which implies that $\dot \theta_i$ must satisfy $|\dot \theta_i| \le \dot \theta_{\text{max}}$ at any iteration round where $\dot \theta_i$ is the $i$th element of $\dot \theta_{ri}$. Since the reference angular velocity $\dot \theta_{r,i}$ is planned in advance, it must satisfy $|\dot \theta_{r,i}| \le \dot \theta_{r,\text{max}}$. Besides, there is a natural requirement, i.e., $\dot \theta_{r,\text{max}} \le \dot \theta_{\text{max}}$.
\subsection{AMCILC design}
\begin{figure}[!tp]
\centering
\includegraphics[width=1\linewidth]{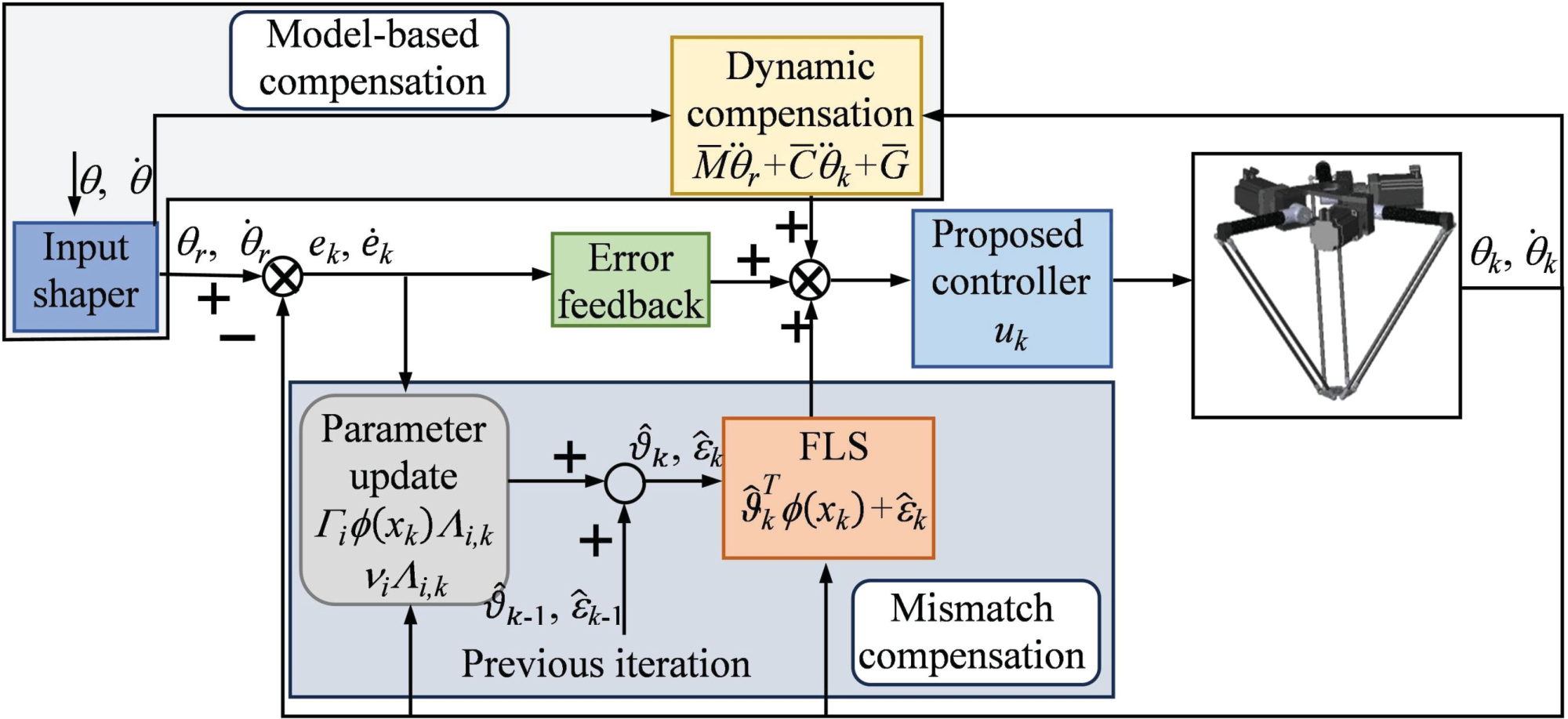}
\caption{Block diagram of the designed control system.}
\label{Block diagram}
\end{figure}

We now introduce the control approach based on concepts from ILC \new{and adaptive control}, and prove the convergence of the proposed approach. \new{The main framework of the proposed control strategy is depicted in Fig. \ref{Block diagram}. The reference trajectory is modified via the designed optimal IS to achieve residual vibration suppression. The proposed controller consists of three parts: dynamic compensation based on the established dynamic model,  feedback coming from the tracking errors, and model mismatch compensation achieved by FLS.}

First of all, by taking the first derivative of \eqref{auxiliary variable} with respect to time and combining \eqref{fuzzy dynamics}, one has
\begin{align}
\label{derivative auxiliary}
    \dot \eta_k=&\ddot e_k+\sigma \dot e_k\nonumber\\
    =&\ddot \theta_k-\ddot\theta_r+\sigma \dot e_k\nonumber\\
    =&\bar M_{r,k}(\theta_k)^{-1}(u_k-\bar C_{r,k}(\theta_k,\dot\theta_k)\dot\theta_k-\bar G_{r,k}(\theta_k)\nonumber\\
    &+\vartheta^T\phi(x_k)+\varepsilon)-\ddot\theta_r+\sigma \dot e_k.
\end{align}
Assumption 1: The initial conditions satisfy the following relationships, i.e., $\theta_k(0)=\theta_r(0)=0$ and $\dot \theta_k(0)=\dot \theta_r(0)=0$, which means $\eta_k(0)=0$.

Then, the proposed AMCILC at the $k$th iteration will be constructed as
\begin{align}
\label{input}
    u_k=&\bar C_r(\theta_k,\dot\theta_k)\dot\theta_k+\bar G(\theta_k)-\hat \vartheta^T_k\phi(x_k)\nonumber\\
    &+\bar M(\theta_k)(\ddot \theta_r-\sigma \dot e_k-k\eta_k)-\hat{\varepsilon}_k, 
\end{align}

where \new{$\hat \vartheta_k$ and $\hat{\varepsilon}_k$ are the estimate of $\vartheta$ and $\varepsilon$ at $k$th iteration, respectively.} The \new{adaptive} parameter update laws are designed as:
\begin{align}
    \hat \vartheta_{i,k}=&\text{sat}(\bar \vartheta_{i,k}),\quad \hat \vartheta_{i,0}=0,\\
    \bar \vartheta_{i,k}=&\hat \vartheta_{i,k-1}+\Gamma_i\phi(x_{k})\Lambda_{i,k},\\
    \hat{\varepsilon}_{i,k}=&\hat{\varepsilon}_{i,k-1}+\nu_i\Lambda_{i,k},\quad \hat \varepsilon_{i,0}=0,\\
    \Lambda_{i,k}=&\Psi_k^T m_{:,i,k},\\
    \Psi_k=&\left[ \frac{\eta_{1,k}}{\cos^2(\frac{\pi\eta_{1,k}^2}{2v_c^2})},\frac{\eta_{2,k}}{\cos^2(\frac{\pi\eta_{2,k}^2}{2v_c^2})},\frac{\eta_{3,k}}{\cos^2(\frac{\pi\eta_{3,k}^2}{2v_c^2})} \right]^T,
\end{align}
where $m_{:,i,k}$ denotes the $i$th column of the inverse of the inertia matrix $\bar M_{r,k}(\theta_k)$. \new{$\hat \vartheta_{i,k}$, $\hat{\varepsilon}_{i,k}$ and $\eta_{i,k}$ denote respectively $i$th element of $\hat \vartheta_k$, $\hat{\varepsilon}_k$ and $\eta_k$, $i=1,2,3$.} $\Gamma_i \in \Re^{l\times l}$ is the adjustable positive diagonal matrix. $\nu_i$ is a positive constant. Assuming the upper and lower boundedness of $\vartheta_{i,k}$ are $\vartheta_{i,\text{max}}$ and $\vartheta_{i,\text{min}}$. $k=\text{diag}(k_1,k_2,k_3)\in\Re^{3\times3}$ is a positive diagonal matrix.

Substituting \eqref{input} into \eqref{derivative auxiliary}, one can have
\begin{align}
        \dot \eta_{k}=&-k\eta_k+\bar M(\theta_k)^{-1}(\tilde{\vartheta}_k^T\phi(x_k)+\varepsilon-\hat \varepsilon_k)\nonumber\\
        =&-k\eta_k+\chi_k,
\end{align}
where $\chi_k=[\chi_{1,k},\chi_{2,k},\chi_{3,k}]^T \in \Re^{3\times1}$ with $\chi_{i,k}=m_{i,1,k}(\tilde{\vartheta}_{1,k}^T\phi(x_k)+\varepsilon_{1}-\hat \varepsilon_{1,k})+m_{i,2,k}(\tilde{\vartheta}_{2,k}^T\phi(x)+\varepsilon_2-\hat \varepsilon_{2,k})+m_{i,3,k}(\tilde{\vartheta}_{3,k}^T\phi(x)+\varepsilon_3-\hat \varepsilon_{3,k})$ in which $m_{i,j,k}$ denotes the element in the $i$th row and $j$th column of the matrix $\bar{M}_{r,k}^{-1}(\theta_k)$ at the $k$th iteration. $\tilde \vartheta_{i,k}=\vartheta_i-\hat \vartheta_{i,k}$ is the estimation error of the weight vector of the fuzzy logic structure at $k$th iteration.

We can now proceed to proving the convergence of the controller under assumption 1. Before that, the following Lemma that will be used in the convergence proof is introduced.

\textit{Lemma 2} \cite{sun2023repetitive}: For $\forall \vartheta_{i,k}\in [\vartheta_{i,\text{min}},\vartheta_{i,\text{max}}]$, the following inequality holds
\begin{equation}
    (\vartheta_{i}-\text{sat}(\bar \vartheta_{i,k}))^T\Gamma_i^{-1}(\bar \vartheta_{i}-\text{sat}(\bar \vartheta_{i,k})\le 0.
\end{equation}

\textit{Theorem 1}: For a dynamic system given by \eqref{rigid body} with output constraints and satisfying assumption 1, the proposed AMCILC can guarantee that tracking errors asymptotically converge to zero when the iteration number $k$ tends to infinity, i.e., $\lim_{k \to \infty } e_k=0,\quad\lim_{k \to \infty } \dot e_k=0$, in which the noise is assumed to be i.i.d. with zero mean. Besides, the output of the system will never exceed the predefined ranges at any iteration.

\textit{Proof}: \new{The proof of the \textit{Theorem 1} is motivated by \cite{xu2002composite}, and it is divided into three parts, i.e., the negativity of the difference of BCEF, the boundedness of $E_0$ and the convergence of the tracking errors. We provide the full derivation in Appendix B.}
\section{Numerical study}

\begin{table}
\begin{center}
\caption{Parameters of the Delta robot.}
\label{tab1}
\begin{tabular*}{\linewidth}{llll}
\toprule
Symbol & Definition & Value & Unit \\ 
\midrule
$l_1$     &     Length of upper arms       &   0.375    &   m   \\
$l_2$       &    Length of lower arms        &   0.95    &   m   \\
$D_1$       &      The outer diameter of upper arms      &    0.058   &   m   \\
$d_1$       &      The internal diameter of upper arms      &    0.048   &   m   \\
$D_2$       &      The outer diameter of lower arms      &   0.016    &   m   \\
$d_2$       &      The internal diameter of lower arms      &    0.012   &   m   \\
$e_a$       &      The radius of the fixed base      &    0.164   &   m   \\
$e_b$       &      The radius of the moving platform      &   0.051    &    m  \\
$E_r$       &       The elastic modulus of all arms    &   71    &   GPa   \\
$\rho_r$       &      Density of upper and lower arms      &    2770   &   Kg/$\text{m}^3$   \\
$\nu_r$       &       Poison's ratio     &    0.3   &      \\
$m_p$       &       Mass of the moving platform     &    0.676   &    Kg  \\
$m_{lump}$       &      The lumped mass      &    0.157   &   Kg   \\ 
$I_{p,x}$       &      Moment of inertia of the MP in $x$-axis     &   2.25    &    Kg$\times\text{mm}^2$   \\
$I_{p,y}$       &      Moment of inertia of the MP in $x$-axis     &   2.25    &    Kg$\times\text{mm}^2$   \\
$I_{p,z}$       &      Moment of inertia of the MP in $x$-axis     &   4.39    &    Kg$\times\text{mm}^2$   \\
$n_{gear}$     & Gear ratio & 15& \\
\bottomrule
\end{tabular*}
\end{center}
\end{table}

\begin{figure}[!t]
\centering
\includegraphics[width=0.7\linewidth]{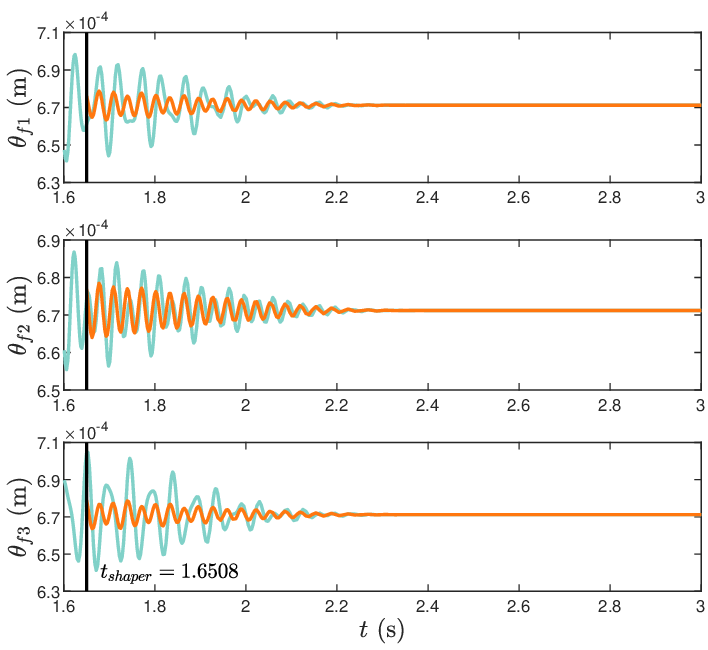}
\caption{The residual vibration of AMCILC (indigo solid lines) and IS-AMCILC (orange solid lines) under the theoretical model.}
\label{defor}
\end{figure}

In order to verify the effectiveness of the proposed IS-AMCILC, we conduct a series of simulations by section A: mathematical model, and section B: high-fidelity Simscape model in which the flexibility of the links, PMSM can be simulated with excellent quality. Since stiffness of the servo system can not be taken into account in the Simscape model, case 1 in section A is primarily used to verify the effective of the designed optimal IS. In section B, we validate the effectiveness of the IS-AMCILC by comparing it to two other controllers: PID-type iterative learning controller (PIDILC) and adaptive fuzzy controller (AFC). Parameters of the robot system are listed in Table \ref{tab1}.
\begin{figure}[!t]
\centering
\includegraphics[width=0.7\linewidth]{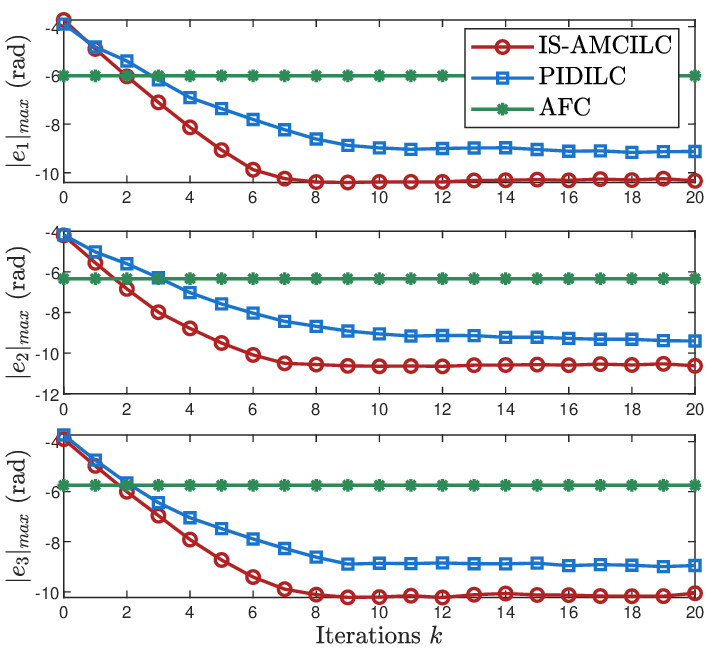}
\caption{Log-scalar maximum angular position tracking error with IS-AMCILC, PIDILC AND AFC.}
\label{fig7}
\end{figure}
\begin{figure}[!t]
\centering
\includegraphics[width=0.7\linewidth]{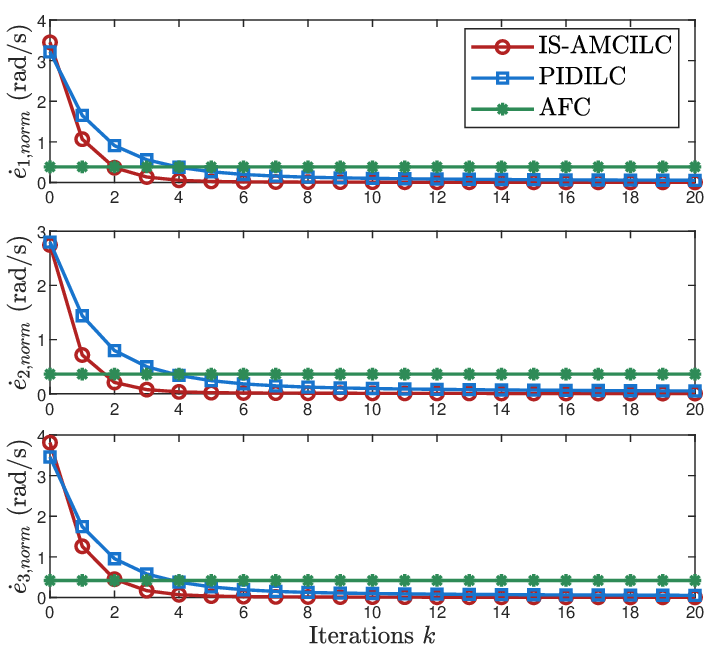}
\caption{The 2-norm of the angular velocity tracking error with IS-AMCILC, PIDILC AND AFC.}
\label{fig8}
\end{figure}
\begin{figure}[!t]
\centering
\includegraphics[width=0.7\linewidth]{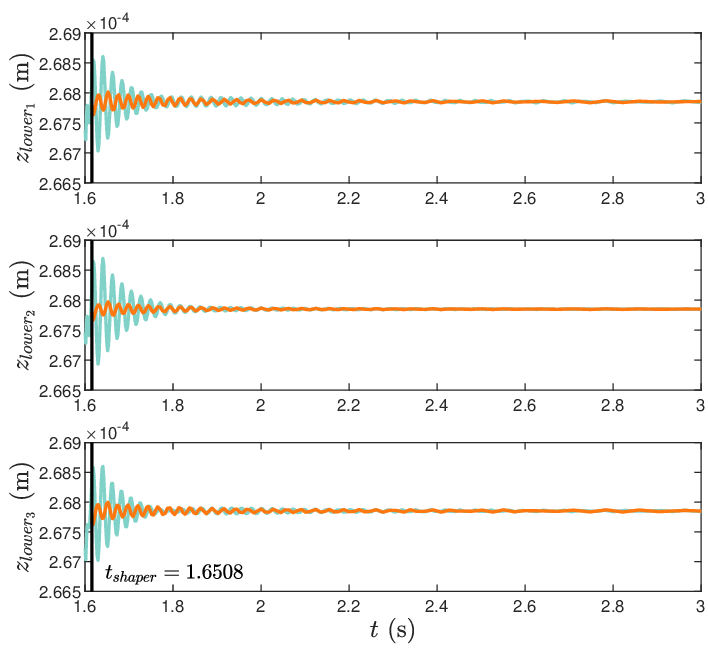}
\caption{The residual vibration of AMCILC (indigo solid lines) and IS-AMCILC (orange solid lines) under the Simscape model.}
\label{fig9}
\end{figure}
\subsection{Case 1-Mathematical model}
Since the stiffness of the servo system cannot be considered in the Simscape model, we conduct a mathematical simulation firstly. The performance of residual vibration suppression is illustrated in Fig. \ref{defor}, where $t_{shaper}$ is the end time of the trajectory after introducing IS. It can be found that there is a significant reduction of the residual vibration by introducing the designed optimal input shaper. Besides, the parameters of the proposed IS-AMCILC in this simulation will be set as follows.

\textit{IS-AMCILC}: For the proposed IS-AMCILC, the control parameters are selected as: $\sigma=1$, $k=I_3$, $v_c=0.1$, $\Gamma_1=\Gamma_2=\Gamma_3=E_9$, $\nu_1=\nu_2=\nu_3=0.01$, where $I_n$ denotes the $n$-order identity matrix. The membership functions of the FLS are designed as

\begin{align}
    \mu_{F_{i}^1}(x_i)=&\frac{1}{1+e^{5(x_i-\kappa_{i,1})}},\mu_{F_{i}^9}(x_i)=\frac{1}{1+e^{5(x_i-\kappa_{i,9})}},\nonumber\\
    \mu_{F_{i}^j}(x_i)=&e^{-\frac{(x_i-\kappa_{i,j})^2}{\psi^2}},j=2,3,\cdots 8,\nonumber
\end{align}
where $x=[\theta^T,\dot\theta^T]^T$. The parameter $\psi$ is set as $\psi=\sqrt{2}$. $\kappa_{i,j}$ denotes the $j$th element of $\kappa_{i}$ and $\kappa_{i}$ is chosen as 
\begin{align}
    \kappa_{1}=&\left\{-0.3,-0.25,-0.2,-0.15,-0.1,\right.\nonumber\\ &\left.-0.05,0,-0.01,0.05\right\},\nonumber\\
   \kappa_{2}=& \left\{ -0.2,-0.15,-0.1,-0.05,-0.03,\right.\nonumber\\ &\left.-0.01,0,-0.01,0.05 \right\},\nonumber\\
   \kappa_{3}=&\left\{ -0.15,-0.05,0,0.05,0.1,0.15,0.2,0.25,0.3 \right\},\nonumber\\
   \kappa_{4}=&\left\{ -1,-0.7,-0.4,-0.1,0,0.1,0.4,0.7,1 \right\},\nonumber\\
   \kappa_{6}=&\kappa_{5}=\kappa_{4}.\nonumber
\end{align}

\begin{figure}[!t]
  \centering
  \subfloat[]{\includegraphics[width=0.45\linewidth]{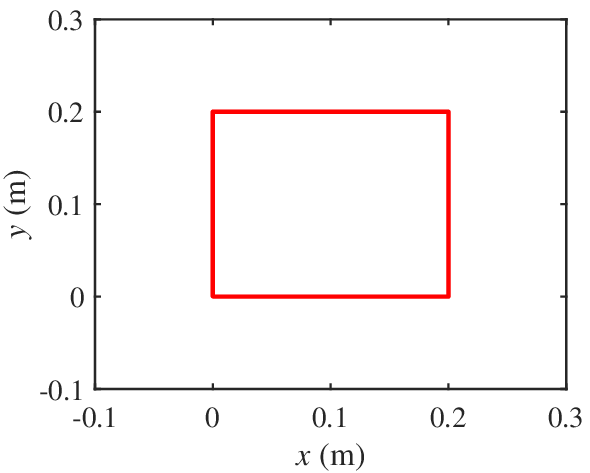}\label{fig:image1}}
  \hfil
  \subfloat[]{\includegraphics[width=0.45\linewidth]{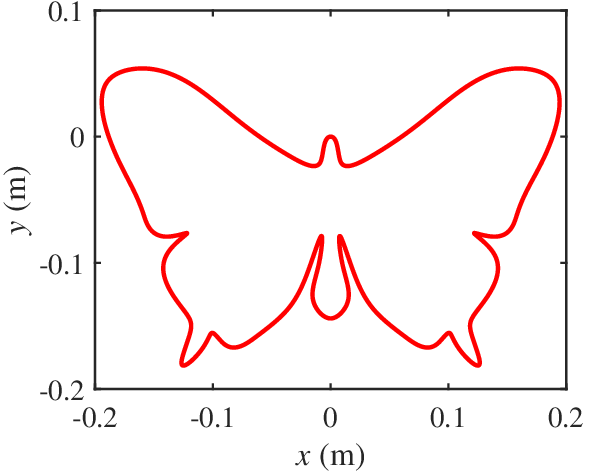}\label{fig:image2}}
  \caption{Two extra trajectories in the $xy$ plane with $z=-0.8151$m; (a) the square trajectory; (b) the butterfly trajectory.}
  \label{fig:two_images}
\end{figure}
\subsection{Case 2-High fidelity model}
In order to further verify the effectiveness of the proposed IS-AMCILC, we establish a multi-physical domain coupling model by Simulink/Simscape. This model contains two parts, i.e., the PMSM and the robot mechanism, where the robot mechanism consists of three rigid-body upper arms and six flexible lower arms. Since the flexibility of the actuation joints is hard to be considered in this Simscape model, only the flexibility of the lower arms will be taken into account. Besides, the PMSM can be chosen from the Simscape built-In libraries. The selection criteria for the PMSMs is based on the output torque requirements of the robots. In this paper, without loss of generality, the PMSMs are selected to be Siemens 1FK7086-4SF7. The mechanical parts and PMSMs are connected by a ''Rotational Multibody Interface'' block. \new{We set the rotor damping to be 0.05 Nm/(rad/s) for three motors in Simscape and treat them as unknown parameters.}

In addition, we introduce two additional control strategies for comparison and the parameters of the three controllers will be set as follows. 

\textit{IS-AMCILC}: For the proposed AMCILC, the control parameters are selected as: $\sigma=1$, $k=\text{diag}(15,15,15)$, $v_c=0.1$, $\Gamma_1=\Gamma_2=\Gamma_3=I_9$, $\nu_1=\nu_2=\nu_3=0.01$. The fuzzy membership function and other parameters are set as identical with that in the theoretical simulation.

\textit{PIDILC}: We use a \revise{PIDILC controller introduced in \cite{arimoto1986mathematical} as comparison. PIDILC is a feedforward memory-based ILC method that enhances traditional PID control by iteratively incorporating previous control inputs.} Firstly, we implement an initial PID controller  to generate the necessary data. 
\begin{equation}
    u(t)=-k_{p0}e(t)-k_{d0}\dot e(t)-k_{i0}\int_{0}^{t}e(\varrho)d\varrho,
\end{equation}
where $k_{p0}, k_{i0}, k_{d0} \in\Re^{3\times3}$ are the control gains, and will be set as: $k_{p0}=20I_3$, $k_{i0}=20I_3$, $k_{d0}=10I_3$.

Then, we introduce a PIDILC from \cite{arimoto1986mathematical} as comparison,
\begin{align}
    u_k=u_{k-1}-k_{p}e(t)-k_{d}\dot e(t)-k_{i}\int_{0}^{t}e(\varrho)d\varrho,
\end{align}
where the control gains are set as: $k_{p}=I_3$, $k_{d}=I_3$ and $k_{i}=I_3$.

\textit{AFC}: \revise{AFC is an effective way to address controller design of nonlinear systems with model mismatch, in which the FLS is used to learn the unknown mismatch \cite{jihong2007robust}.} An AFC strategy is employed here.
\begin{align}
    u=&\bar C_r(\theta_{ri},\dot\theta)\dot\theta_{ri}+\bar G_r(\theta_{ri})-\hat \vartheta^T\phi(x),\nonumber\\
    &+\bar M_r(\theta_{ri})(\ddot \theta_r-\sigma \dot e-k\eta)-\hat{\varepsilon},   
\end{align}

with the update laws 
\begin{align*}
    \dot {\hat \vartheta}_{i}=\Gamma_i\phi(x)\Lambda_{i},\ 
    \dot {\hat{\varepsilon}}_{i}=\nu_i\Lambda_{i},
\end{align*}
where the fuzzy membership function is the same with IS-AMCILC. and other control gains are selected as: $\sigma=1$, $k=10I_3$, $\Gamma_1=\Gamma_2=\Gamma_3=20I_9$, $\nu_1=\nu_2=\nu_3=0.1I_3$ and $v_c=0.1$. \new{Note that, the AFC is related to the AMCILC in (22) but the parameter update laws do not use the iterative learning structure.}
\begin{figure}[!t]
\centering
\includegraphics[width=0.75\linewidth]{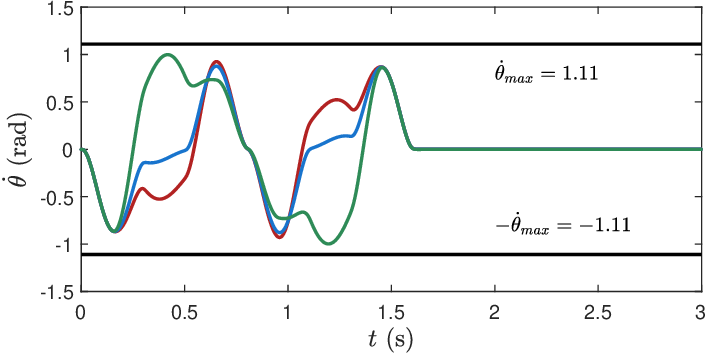}
\caption{The output angular velocity at 3th iteration: $\theta_1$ (red solid line); $\theta_2$ (blue solid line); $\theta_3$ (green solid line).}
\label{velocity}
\end{figure}
\subsection{Results}
Fig. \ref{fig7} shows the log-scale angular position tracking performance of three controllers. Both the IS-AMCILC and PIDILC can drive the angular position to the reference trajectory, i.e., the tracking errors will converge to zero along the iteration axis, and in which the $|e_i|_{max}$ denotes the maximum absolute angular position tracking error of $i$th input joints. However, \revise{thanks to FLS, the controller can achieve good tracking performance even in the presence of mismatches.} It can be obviously found that the proposed IS-AMCILC can converge to zeros quicker compared to the PIDILC. Besides, since the initial tracking error of PIDILC is generated by a PID controller, the initial tracking error for these two controllers is different. But we have tried to make them as close as possible. Fig. \ref{fig8} illustrates that the angular velocity tracking errors will also decrease along the iteration axis where $\dot e_{i,norm}$ denotes the 2-norm of the angular velocity tracking error vector of $i$th input joints, but the angular velocity tracking performance of the proposed IS-AMCILC is obviously much better than the PIDILC. For algorithm 3, since it is not an iterative learning algorithm, the tracking errors will never change along the iteration axis. Although it can achieve a better performance than the proposed IS-AMCILC under the initial iteration, the proposed IS-AMCILC can achieve a better performance from 2nd iteration on. 

The performance of the designed optimal input shaper can be seen from Fig. \ref{fig9} in which $z_{lower_i}$ denotes the residual vibration in the $z$-axis of the points connecting the $i$th lower arms and the MP. By introducing the designed optimal input shaper, the residual vibration can be suppressed significantly. Besides, the 3rd output angular velocity is shown in Fig. \ref{velocity}, which demonstrates that the output angular velocity will never exceed the constraints.

In order to further verify the performance of the proposed AMCILC, we conduct a series of  simulations with two extra planar trajectories: the square trajectory represented by $SE$, and the butterfly trajectory represented by $BY$. The tracking performance of these two trajectories is evaluated in Tables \ref{tab2} and \ref{tab3}.
Both trajectories start with larger errors without proposed method, but following 20 learning iterations improve significantly. The angular position errors of the BY trajectory are initially larger for IS-AMCILC, but improve significantly following 20 ILC iterations, and become 40-50 fold smaller than the PIDILC error, especially for joints 2 and 3.

\begin{table}
\centering
\caption{Angular position tracking performance comparison.}
\label{tab2}
\begin{tabular}{ m{1.5cm}  m{0.6cm} m{0.4cm}  m{0.4cm}  m{0.3cm} m{0.3cm}  m{0.3cm} m{0.3cm} }
\hline
                      &        & \multicolumn{2}{c}{IS-AMCILC} & \multicolumn{2}{c}{PIDILC} & \multicolumn{2}{c}{AFC} \\ \cmidrule(lr){3-4} \cmidrule(lr){5-6} \cmidrule(lr){7-8} 
                      &        & $SE$     & $BY$    & $SE$     & $BY$    & $SE$     & $BY$    \\ \hline
\multirow{3}{*}{0\textit{th} [$10^{-2}$rad]}   & $|e_1|_{max}$ & 1.78         & 3.38        & 2.31         & 2.45        & 0.44         & 0.44        \\
                                    & $|e_2|_{max}$ & 2.06         & 3.21        & 2.92         & 2.15        & 0.70         & 0.35        \\
                                    & $|e_3|_{max}$ & 3.11         & 2.27        & 3.45         & 1.86        & 0.69         & 0.26        \\ \hline
\multirow{3}{*}{20\textit{th} [$10^{-3}$rad]}  & $|e_1|_{max}$ & 0.069         &0.11        & 0.28         & 0.13        & \textbackslash         & \textbackslash        \\
                                    & $|e_2|_{max}$ & 0.055         & 0.053        & 0.30         & 0.18        & \textbackslash         & \textbackslash        \\
                                    & $|e_3|_{max}$ & 0.074         & 0.051        & 0.30         & 0.13        & \textbackslash         & \textbackslash        \\ \hline
 
\end{tabular}
\end{table}

\begin{table}
\centering
\caption{Angular velocity tracking performance comparison.}
\label{tab3}
\begin{tabular}{ m{1.6cm}  m{0.6cm} m{0.4cm}  m{0.4cm}  m{0.3cm} m{0.3cm}  m{0.3cm} m{0.3cm} }
\hline
                      &        & \multicolumn{2}{c}{IS-AMCILC} & \multicolumn{2}{c}{PIDILC} & \multicolumn{2}{c}{AFC} \\ \cmidrule(lr){3-4} \cmidrule(lr){5-6} \cmidrule(lr){7-8} 
                      &        & $SE$      & $BY$    & $SE$      & $BY$    & $SE$      & $BY$    \\ \hline
\multirow{3}{*}{0\textit{th} [rad/s]}   & $|\dot e_1|_{norm}$ & 0.08         & 0.1        & 0.2         & 0.19        & 0.03         & 0.02        \\
                                    & $|\dot e_2|_{norm}$ & 0.09         & 0.1        & 0.2         & 0.13        & 0.03         & 0.03        \\
                                    & $|\dot e_3|_{norm}$ & 0.1         & 0.1        & 0.2         & 0.18        & 0.03         & 0.02        \\ \hline
\multirow{3}{*}{20\textit{th} [$10^{-2}$rad/s]}  & $|\dot e_1|_{norm}$ & 0.2         &0.1        & 1.1         & 0.72        & \textbackslash         & \textbackslash        \\
                                    & $|\dot e_2|_{norm}$ & 0.6         & 0.1        & 2.0         & 0.83        & \textbackslash         & \textbackslash        \\
                                    & $|\dot e_3|_{norm}$ & 0.6         & 0.1       & 2.0         & 0.87        & \textbackslash         & \textbackslash        \\ \hline
 
\end{tabular}
\end{table}

\section{Conclusion}
This article developed an input shaping techniques based AMCILC strategy for Delta robot with angular velocity constraints to address the problem of trajectory tracking and residual vibration suppression simultaneously. An optimal input shaper was designed to achieve the optimal residual vibration suppression in the whole workspace. An AMCILC was designed to achieve high performance trajectory tracking, in which the FLS was introduced to approximate the model mismatch. A BLF was proposed to satisfy the angular velocity constraints. In addition, A BECF was utilized to prove the convergence of the tracking errors along the iteration axis. Moreover, the simulation results have illustrated that the proposed control strategy was effective for the Delta robot system.
{\appendix[]
\subsection{Appendix A}
\new{
The closed-loop structure of Delta robots implies that the deformations of the end point of each lower arm that connects with the MP and the displacements and rotations of the MP are not independent. The deformation constraints are shown in Fig. \ref{fig_2}. The closed-loop vector relationship before the occurrence of deformation is given as
\begin{equation}
    r_a=P+r_b \label{fr},
\end{equation}
where $r_a\in \Re^{3\times1}$ denotes the vector of one kinematic chain, and $P\in \Re^{3\times1}$ is the vector of the center point of the MP. $r_b\in \Re^{3\times1}$ is the vector pointing from the center point of the MP to the connecting point. All vectors are measured in global coordinate system.
}
\new{
Then, after the presence of deformation, the closed-loop vector is rewritten as
\begin{equation}
    r_a+\delta_a=P+\xi+\delta Rr_b \label{rr},
\end{equation}
where $\delta_a\in \Re^{3\times1}$ is the deformation of the connection points. $\xi\in \Re^{3\times1}$ is the small displacement of the MP. $\delta R\in \Re^{3\times3}$ is the transformation matrix from $O_{\Tilde{p}}-x_{\Tilde{p}}y_{\Tilde{p}}z_{\Tilde{p}}$ to $O-xyz$ caused by the small rotations. Without loss of generality, the $Z\rule[0.1cm]{2mm}{1pt}Y\rule[0.1cm]{2mm}{1pt}X$ Euler angle $\delta_p=(\delta_x,\delta_y,\delta_z)^T\in \Re^{3\times1}$ is utilized to describe this rotations. Therefore, the transformation matrix $\delta R$ is expressed as
\begin{equation}
\delta R=\begin{bmatrix}
c\delta_xc\delta_y & c\delta_xs\delta_ys\delta_z-s\delta_xc\delta_z & c\delta_xs\delta_yc\delta_z+s\delta_xs\delta_z \\
 s\delta_xc\delta_y& s\delta_xs\delta_ys\delta_z+c\delta_xc\delta_z & s\delta_xs\delta_yc\delta_z-c\delta_xs\delta_z \\
-s\delta_y & c\delta_ys\delta_z & c\delta_yc\delta_z
\end{bmatrix}, \label{dr}
\end{equation}
where $c$ and $s$ denote $\cos$ and $\sin$, respectively. Since the rotation is extremely tiny, the following relations can be reasonably deduced, which is $s(\ast)\approx \ast$
and $c(\ast)\approx 1$. Hence, \eqref{dr} can be simplified as 
\begin{equation}
    \delta R=\begin{bmatrix}
1 & -\delta_x & \delta_y \\
 \delta_x& 1 & -\delta_z \\
-\delta_y & \delta_z & 1
\end{bmatrix}. \label{RR}
\end{equation}
}

\new{
Combining \eqref{fr} with \eqref{rr} and utilising \eqref{RR}, we have the following relation
\begin{equation}
\begin{split}
    \delta_a=\xi+\delta Rr_b-r_b
    =\xi-[r_b\times]I_\delta\delta_p=T_ld_p
\end{split},
\end{equation}
where $[r_b\times]$ denotes the skew-symmetric matrix of $r_b$, and $I_\delta$ represents a 3-order anti-diagonal identity matrix. $d_p=[\xi, \delta_p]^T\in \Re^{6\times1}$, and $T_l \in \Re^{3\times6}$ denotes the deformation compatibility matrix for one lower arm link.}

\new{Based on the above relationship for each lower arm link (Delta robots have 6 lower arm links), we can deduce \eqref{compatibility} easily.}
\subsection{Appendix B}
This is the proof of \textit{Theorem 1}. First of all, \new{considering the tracking error, the FLS parameter error and approximation error,} we develop a BCEF as 
\begin{equation}
\label{BCEF}
    E_k(t)=V_{\eta,k}(t)+V_{\vartheta,k}(t)+V_{\varepsilon,k}(t),
\end{equation}
where
\begin{align}
    V_{\eta,k}(t)=&\sum_{i=1}^{3} \frac{v_c^2}{\pi}\tan(\frac{\pi\eta_{i,k}^2(t)}{2v_c^2}),\\
    V_{\vartheta,k}(t)=&\sum_{i=1}^{3}\frac{1}{2}\int_{0}^{t}\tilde{\vartheta}_{i,k}^T(\varrho)\Gamma_i^{-1}\tilde{\vartheta}_{i,k}(\varrho)d\varrho,\\
    V_{\varepsilon,k}(t)=&\sum_{i=1}^{3} \frac{1}{2\upsilon_i}\int_{0}^{t}(\bar\varepsilon_i-\hat\varepsilon_{i,k}(\varrho))^2d\varrho.
\end{align}

\textit{Part A}-The Non-increasing Property of BCEF. \new{In order to verify the non-increasing property of (38), we first derive $V_{\eta,k}(t)-V_{\eta,k-1}(t)$, $V_{\vartheta,k}(t)-V_{\vartheta,k-1}(t)$ and $V_{\varepsilon,k}(t)-V_{\varepsilon,k-1}(t)$ as follows.}
\begin{align}
\label{term1}
&V_{\eta,k}(t)-V_{\eta,k-1}(t)\nonumber\\
=&\sum_{i=1}^{3} \left( \frac{v_c^2}{\pi}\tan(\frac{\pi\eta_{i,k}^2}{2v_c^2})-\frac{v_c^2}{\pi}\tan(\frac{\pi\eta_{i,k-1}^2}{2v_c^2}) \right)\nonumber\\
\le&\sum_{i=1}^{3}\int_{0}^{t}\left( \Lambda_{i,k} \tilde{\vartheta}_{i,k}^T\phi(x_k)+\Lambda_{i,k}\left(\bar{\varepsilon}_i-\hat{\varepsilon}_{i,k} \right) \right)d\varrho\nonumber\\
&-\Psi_k^Tk\eta_k-\sum_{i=1}^{3}\frac{v_c^2}{\pi}\tan(\frac{\pi\eta_{i,k-1}^2}{2v_c^2}),
\end{align}
where the assumption 1 has been used here. Then, for the second term, we have
\begin{align}
&V_{\vartheta,k}(t)-V_{\vartheta,k-1}(t)\nonumber\\
=&\sum_{i=1}^{3}\frac{1}{2}\int_{0}^{t}(\vartheta_i(\varrho)-\hat\vartheta_{i,k}(\varrho))^T\Gamma_i^{-1}(\vartheta_i(\varrho)-\hat\vartheta_{i,k}(\varrho))d\varrho\nonumber\\
&-\sum_{i=1}^{3}\frac{1}{2}\int_{0}^{t}(\vartheta_i(\varrho)-\hat\vartheta_{i,k-1}(\varrho))^T\Gamma_i^{-1}\nonumber\\&\times(\vartheta_i(\varrho)-\hat\vartheta_{i,k-1}(\varrho))d\varrho\nonumber\\
=&\sum_{i=1}^{3}\int_{0}^{t}(\vartheta_i(\varrho)-\hat\vartheta_{i,k}(\varrho))^T\Gamma_i^{-1}(\hat\vartheta_{i,k-1}(\varrho)-\hat\vartheta_{i,k}(\varrho))d\varrho\nonumber\\
&-\sum_{i=1}^{3}\frac{1}{2}\int_{0}^{t}(\hat\vartheta_{i,k}(\varrho)-\hat\vartheta_{i,k-1}(\varrho))^T\Gamma_i^{-1}\nonumber\\
&\times(\hat\vartheta_{i,k}(\varrho)-\hat\vartheta_{i,k-1}(\varrho))d\varrho.
\end{align}

By means of Lemma 2, and the adaptive update laws, one can have
\begin{align}
\label{term2}
&V_{\vartheta,k}(t)-V_{\vartheta,k-1}(t)\nonumber\\
=&\sum_{i=1}^{3}\int_{0}^{t}(\vartheta_i(\varrho)-\text{sat}(\bar \vartheta_{i,k}(\varrho)))^T\Gamma_i^{-1}\nonumber\\&\times(\bar \vartheta_{i,k}(\varrho)-\Gamma_i\phi(x)\Lambda_i-\text{sat}(\bar\vartheta_{i,k}(\varrho)))d\varrho\nonumber\\&-\sum_{i=1}^{3}\frac{1}{2}\int_{0}^{t}(\hat\vartheta_{i,k}(\varrho)-\hat\vartheta_{i,k-1}(\varrho))^T\nonumber\\
&\times\Gamma_i^{-1}(\hat\vartheta_{i,k}(\varrho)-\hat\vartheta_{i,k-1}(\varrho))d\varrho\nonumber\\
\le&-\sum_{i=1}^{3}\frac{1}{2}\int_{0}^{t}(\hat\vartheta_{i,k}(\varrho)-\hat\vartheta_{i,k-1}(\varrho))^T\Gamma_i^{-1}\times(\hat\vartheta_{i,k}(\varrho)\nonumber\\
&-\hat\vartheta_{i,k-1}(\varrho))d\varrho-\sum_{i=1}^{3}\int_{0}^{t}\tilde\vartheta_{i,k}^T(\varrho)\phi(x_k)\Lambda_{i,k}d\varrho.
\end{align}

For the last term, it can be rearranged as
\begin{align}
\label{term3}
&V_{\varepsilon,k}(t)-V_{\varepsilon,k-1}(t)\nonumber\\
=&-\sum_{i=1}^{3}\frac{1}{2\nu_i}\int_{0}^{t}(\hat\varepsilon_{i,k}(\varrho)-\hat\varepsilon_{i,k-1}(\varrho))^2d\varrho\nonumber\\
&-\sum_{i=1}^{3}\int_{0}^{t}(\bar\varepsilon_{i}-\hat\varepsilon_{i,k}(\varrho))\Lambda_{i,k}d\varrho.
\end{align}

Consequently, combining \eqref{term1}, \eqref{term2} and \eqref{term3}, one can have
\begin{align}
&E_k(t)-E_{k-1}(t)\nonumber\\
\le&\sum_{i=1}^{3}\int_{0}^{t}\left( \Lambda_{i,k} \tilde{\vartheta}_{i,k}^T\phi(x_k)+\Lambda_{i,k}\left(\bar{\varepsilon}_i-\hat{\varepsilon}_{i,k} \right) \right)d\varrho\nonumber\\&-\Psi_k^Tk\eta_k\nonumber-\sum_{i=1}^{3}\frac{v_c^2}{\pi}\tan(\frac{\pi\eta_{i,k-1}^2}{2v_c^2})\nonumber\\&-\sum_{i=1}^{3}\frac{1}{2}\int_{0}^{t}(\hat\vartheta_{i,k}(\varrho)-\hat\vartheta_{i,k-1}(\varrho))^T\nonumber\\
&\times\Gamma_i^{-1}(\hat\vartheta_{i,k}(\varrho)-\hat\vartheta_{i,k-1}(\varrho))d\varrho-\nonumber\\&\sum_{i=1}^{3}\int_{0}^{t}\tilde\vartheta_{i,k}^T(\varrho)\phi(x_k)\Lambda_{i,k}d\varrho\nonumber\\
&-\sum_{i=1}^{3}\frac{1}{2\nu_i}\int_{0}^{t}(\hat\varepsilon_{i,k}(\varrho)-\hat\varepsilon_{i,k-1}(\varrho))^2d\varrho\nonumber\\
&-\sum_{i=1}^{3}\int_{0}^{t}(\bar\varepsilon_{i}(\varrho)-\hat\varepsilon_{i,k}(\varrho))\Lambda_{i,k}d\varrho\nonumber\\
=&-\Psi_k^Tk\eta_k-\sum_{i=1}^{3}\frac{1}{2\nu_i}\int_{0}^{t}(\hat\varepsilon_{i,k}(\varrho)-\hat\varepsilon_{i,k-1}(\varrho))^2d\varrho\nonumber\\
&-\sum_{i=1}^{3}\frac{v_c^2}{\pi}\tan(\frac{\pi\eta_{i,k-1}^2}{2v_c^2})\nonumber\\&-\sum_{i=1}^{3}\frac{1}{2}\int_{0}^{t}(\hat\vartheta_{i,k}(\varrho)-\hat\vartheta_{i,k-1}(\varrho))^T\nonumber\\
&\times\Gamma_i^{-1}(\hat\vartheta_{i,k}(\varrho)-\hat\vartheta_{i,k-1}(\varrho))d\varrho\nonumber\\
\le&0.
\end{align}
Part B-The Boundedness of $E_0(t)$.
Firstly, according to \eqref{BCEF}, one can have
\begin{align}
\label{e0}
    &E_0(t)\nonumber\\
    =&\sum_{i=1}^{3} \frac{v_c^2}{\pi}\tan(\frac{\pi\eta_{i,0}^2(t)}{2v_c^2})+\sum_{i=1}^{3} \frac{1}{2\upsilon_i}\int_{0}^{t}(\bar\varepsilon_i-\hat\varepsilon_{i,0}(\varrho))^2d\varrho\nonumber\\&+\sum_{i=1}^{3}\frac{1}{2}\int_{0}^{t}\tilde{\vartheta}_{i,0}^T(\varrho)\Gamma_i^{-1}\tilde{\vartheta}_{i,0}(\varrho)d\varrho.
\end{align}

Based on the update laws, the following relationships can be easily obtained
\begin{align}
\label{in1}
    \frac{1}{2}\tilde{\vartheta}_{i,0}^T(t)\Gamma_i^{-1}\tilde{\vartheta}_{i,0}(t)\le&\frac{1}{2}\vartheta_i^T\Gamma_i^{-1}\vartheta_i
    \\&-(\vartheta_i-\hat\vartheta_{i,0}(t))^T\Gamma_i^{-1}\hat\vartheta_{i,0}(t)\nonumber\\
    \label{in2}
    \frac{1}{2\upsilon_i}(\bar\varepsilon_i-\hat\varepsilon_{i,0}(t))^2\le&\frac{1}{2\upsilon_i}\bar\varepsilon_i^2
     -\frac{1}{\upsilon_i}(\bar\varepsilon_i-\hat\varepsilon_{i,0}(t))\hat\varepsilon_{i,0}(t).
\end{align}

Then, by taking the derivative of \eqref{e0} with respect to time and combining \eqref{in1} and \eqref{in2}, we have
\begin{align}
\label{dote}
    &\dot E_0(t)\nonumber\\
    =&\sum_{i=1}^{3} \frac{\eta_{i,0}(t)\dot \eta_{i,0}(t)}{\cos^2(\frac{\pi\eta_{i,0}^2(t)}{2v_c^2})}+\sum_{i=1}^{3}\frac{1}{2}\tilde{\vartheta}_{i,0}^T(t)\Gamma_i^{-1}\tilde{\vartheta}_{i,0}(t)\nonumber\\
    &+\sum_{i=1}^{3} \frac{1}{2\upsilon_i}(\bar\varepsilon_i-\hat\varepsilon_{i,0}(t))^2\nonumber\\
    \le&\sum_{i=1}^{3}\Lambda_{i,0}\left( \tilde{\vartheta}_{i,0}^T\phi(x_0)+\bar\varepsilon_i-\hat{\varepsilon}_{i,0} \right)+\frac{1}{2}\vartheta_i^T\Gamma_i^{-1}\vartheta_i\nonumber\\
    &-(\vartheta_i-\hat\vartheta_{i,0}(t))^T\Gamma_i^{-1}\hat\vartheta_{i,0}(t)+\frac{1}{2\upsilon_i}\bar\varepsilon_i^2\nonumber\\
    &-\frac{1}{\upsilon_i}(\bar\varepsilon_i-\hat\varepsilon_{i,0}(t))\hat\varepsilon_{i,0}(t)-\Psi_0^Tk\eta_0\nonumber\\
    =&\sum_{i=1}^{3}(\vartheta_i-\text{sat}(\bar\vartheta_{i,0}(t)))^T\Gamma_i^{-1}(\bar\vartheta_{i,0}(t)-\text{sat}(\bar\vartheta_{i,0}(t)))\nonumber\\
    &+\frac{1}{2}\vartheta_i^T\Gamma_i^{-1}\vartheta_i+\frac{1}{2\upsilon_i}\bar\varepsilon_i^2-\Psi_0^Tk\eta_0\nonumber\\
    \le&-\Psi_0^Tk\eta_0+\sum_{i=1}^{3}\frac{1}{2\upsilon_i}\bar\varepsilon_i^2+\sum_{i=1}^{3}\frac{1}{2}\vartheta_i^T\Gamma_i^{-1}\vartheta_i.
\end{align}

Therefore, in terms of \eqref{dote}, it can be found that $\dot E_0(t)$ is bounded, which implies that there exists a positive constant $c_0$ satisfying $|\dot E_0(t)|\le c_0$, and we have
\begin{equation}
    E_0(T)\le E_0(0)+\int_{0}^{T}|E_0(\varrho)|d\varrho<E_0(0)+c_0T<\infty,
\end{equation}
where $T$ is the duration of the trajectory. Finally, the boundedness of $E_0(t)$ can be proven.

Part C-The Convergence of Tracking Errors.
\begin{align}
\label{limit}
E_{k}(t)=&E_0(t)+\sum_{j=1}^{k}\Delta E_j(t)\nonumber\\
\le&E_0(t)-\sum_{j=1}^{k}\sum_{i=1}^{3}\frac{v_c^2}{\pi}\tan(\frac{\pi\eta_{i,j-1}^2}{2v_c^2}).
\end{align}

By taking the limit on both sides of \eqref{limit}, one can has
\begin{align}
\lim_{k \to \infty } E_{k}(t)\le E_0(t)-\lim_{k \to \infty } \sum_{j=1}^{k}\sum_{i=1}^{3}\frac{v_c^2}{\pi}\tan(\frac{\pi\eta_{i,j-1}^2}{2v_c^2}).
\end{align}

Since $E_0(t)$ is bounded and $E_k(t)\ge 0$, we have
\begin{align}
\lim_{k \to \infty } \sum_{j=1}^{k}\sum_{i=1}^{3}\frac{v_c^2}{\pi}\tan(\frac{\pi\eta_{i,j-1}^2}{2v_c^2})\le E_0(t).
\end{align}

Finally, by means of the convergence property of the sum of series \cite{liu2019rbfnn}, one can have

\begin{align}
\lim_{k \to \infty } \sum_{i=1}^{3}\frac{v_c^2}{\pi}\tan(\frac{\pi\eta_{i,k-1}^2}{2v_c^2})=0.
\end{align}

Therefore, the relationship $\lim_{k \to \infty } \eta_{i,k-1}=0$ with $i=1,2,3$ holds, which implies $e_k\to 0$ and $\dot e_k\to 0$ as $k \to \infty$.

 Since $V_{\eta,k}(t)$ is bounded, the auxiliary variable $\eta_i$ satisfies $|\eta_i|<v_c$ at any iteration. Then, based on $\eta_i=e_i+\sigma \dot e_i$, we can find a bounded function $f_v(t)$, which satisfies $0<|f_v(t)|\le v_c$ and $|\dot e_{i,k}|<|f_v(t)|$. Besides, due to $\dot e_{i,k}=\dot \theta_{i,k}-\dot\theta_{i,r}$, $|\dot \theta_{i,r}|\le \dot \theta_{r,max}$ and $\dot\theta_{r,max} < \dot \theta_{max}$, there must exist a proper $v_c$ satisfying $0<|f_v(t)|\le v_c <\dot \theta_{max}-\dot\theta_{r,max}$. Then, the relationship $|\dot \theta_{i,k}|\le|\dot e_{i,k}|+|\dot \theta_{i,r}|<|f_v(t)|+\dot \theta_{r,max}<v_c+\dot \theta_{r,max}<\dot \theta_{max}$, which means the constraints of angular velocity can be guaranteed.}



\bibliographystyle{IEEEtran}
\bibliography{IEEEabrv,myrefs}
\vspace{-30 mm}
\begin{IEEEbiography}[{\includegraphics[width=1in,height=1.25in,clip,keepaspectratio]{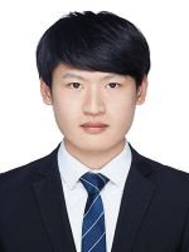}}]{Mingkun Wu} received the B.S. degree in Mechanical Design, Manufacturing, and Automation from Hefei University of Technology, Hefei, China, in 2018 and M.S. degree in Mechanical Engineering from Tianjin University, Tianjin, China, in 2021, respectively. He is currently pursuing Ph.D. degree in Mechanical Engineering with the Institute of Mechanism Theory, Machine Dynamics and Robotics, RWTH Aachen University, Aachen, Germany.

His research interests include parallel robots, iterative learning control, trajectory optimization, vibration suppression.
\end{IEEEbiography}
\vspace{-30 mm}
\begin{IEEEbiography}[{\includegraphics[width=1in,height=1.25in,clip,keepaspectratio]{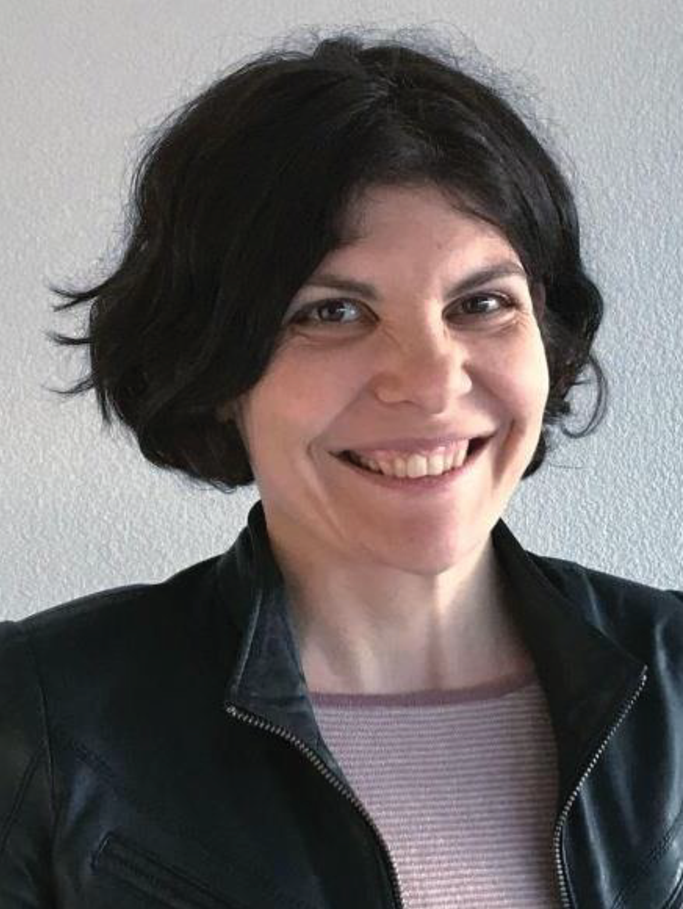}}]{Alisa Rupenyan} (Member, IEEE) received the B.Sc. degree in engineering physics and M.Sc. degree in laser physics from the University of Sofia, Sofia, Bulgaria, in 2004 and 2005, respectively, and the Ph.D. degree from the Department of Physics and Astronomy, Vrije Universiteit Amsterdam, Amsterdam, The Netherlands. She holds the Rieter endowed professorship for Industrial AI at the ZHAW Centre for AI, Zurich University for Applied Sciences, Switzerland. Between 2011 and 2014, she was a Postdoctoral Fellow with ETH Zürich, Zürich, Switzerland, and between 2014 and 2018, she was a Lead Scientist in a robotic start-up. She was a Group Leader in Automation with Inspire AG, the technology transfer unit at ETH Zürich, and a Senior Scientist and PI with the Automatic Control Laboratory, ETH Zurich, between 2018-2023. Her research interests include the intersection between machine learning, control, and optimization for industrial applications and robotics, especially in Bayesian methods applied for optimization and control, as well as learning-based control.
Dr. Rupenyan is an expert for the Swiss Innovation Agency (Innosuisse), a member of several technical committees at IEEE-CSS, IEEE-RAS, and IEEE-IES, and a member of the executive committee at the IFAC Industry Committee.
\end{IEEEbiography}
\vspace{-30 mm}
\begin{IEEEbiography}[{\includegraphics[width=1in,height=1.25in,clip,keepaspectratio]{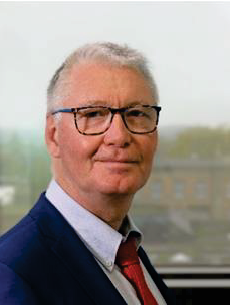}}]{Burkhard Corves} received Ph.D. degree from Department of Mechanism Theory and Machine Dynamics, RWTH Aachen University, Aachen, Germany. Prof. Dr.-Ing. h. c. Burkhard Corves is Director of the Institute of Mechanism Theory, Machine Dynamics and Robotics of RWTH Aachen University, Aachen, Germany since 2000. He is Chairman of the VDI-Committee of the VDI-Scientific Board on “Mechanism and Machine Theory” and Chair of IFToMM MO Germany.
\end{IEEEbiography}
\end{document}